\documentclass[twocolumn, aps, amsmath, amssymb, nofootinbib, superscriptaddress, doublefloatfix, table-of-contents, eqsecnum, pra]{revtex4-2}

\usepackage[pdftex]{graphicx}
\usepackage{mathrsfs}
\usepackage[colorlinks, breaklinks]{hyperref}
\usepackage{amsmath}
\usepackage[english]{babel}
\usepackage{booktabs}
\usepackage{amssymb}
\usepackage{type1cm}
\usepackage{url}
\usepackage[breaklinks]{hyperref}
\usepackage{braket}
\usepackage{algorithm}
\usepackage{algpseudocodex}
\usepackage{multirow}

\usepackage{comment}
\usepackage{graphicx}

\usepackage[caption=false]{subfig} 
\usepackage{ragged2e} 
\DeclareCaptionJustification{justified}{\justifying}

\begin{document}

\title{Proof-of-work consensus by quantum sampling}

\author{Deepesh Singh}
\affiliation{Centre for Quantum Computation \& Communications Technology, School of Mathematics \& Physics, The University of Queensland, St Lucia QLD, Australia}


\author{Gopikrishnan Muraleedharan}\email[]{gopikrishnan.muraleedharan@mq.edu.au}
\affiliation{Center for Engineered Quantum Systems, School of Mathematical and Physical Sciences, Macquarie University, NSW 2109, Australia}

\author{Boxiang Fu}
\affiliation{School of Physics, University of Melbourne, Melbourne, VIC 3010, Australia}


\author{Chen-Mou Cheng}
\affiliation{BTQ Technologies, 16-104 555 Burrard Street, Vancouver BC, V7X 1M8 Canada}

\author{Nicolas Roussy Newton}
\affiliation{BTQ Technologies, 16-104 555 Burrard Street, Vancouver BC, V7X 1M8 Canada}

\author{Peter P. Rohde}
\affiliation{Centre for Quantum Software \& Information (QSI), University of Technology Sydney, NSW, Australia}
\affiliation{Center for Engineered Quantum Systems, School of Mathematical and Physical Sciences, Macquarie University, NSW 2109, Australia}
\affiliation{Hearne Institute for Theoretical Physics, Department of Physics \& Astronomy, Louisiana State University, Baton Rouge LA, United States}


\author{Gavin K. Brennen}
\affiliation{Center for Engineered Quantum Systems, School of Mathematical and Physical Sciences, Macquarie University, NSW 2109, Australia}

\begin{abstract}
Since its advent in 2011, boson sampling has been a preferred candidate for demonstrating quantum advantage because of its simplicity and near-term requirements compared to other quantum algorithms. We propose to use a variant, called coarse-grained boson-sampling (CGBS), as a quantum Proof-of-Work (PoW) scheme for blockchain consensus. The users perform boson sampling using input states that depend on the current block information and commit their samples to the network. Afterwards, CGBS strategies are determined which can be used to both validate samples and reward successful miners. By combining rewards for miners committing honest samples together with penalties for miners committing dishonest samples, a Nash equilibrium is found that incentivizes honest nodes. We provide numerical evidence that these validation tests are hard to spoof classically without knowing the binning scheme ahead of time and show the robustness of our protocol to small partial distinguishability of photons. The scheme works for both Fock state boson sampling and Gaussian boson sampling and provides dramatic speedup and energy savings relative to computation by classical hardware. 
\end{abstract}

\maketitle

Blockchain technology relies on the ability of a network of non-cooperating participants to reach consensus on validating and verifying a new set of block-bundled transactions, in a setting without centralized authority. A consensus algorithm is a procedure through which all the peers of the blockchain network reach a common agreement about the present state of the distributed ledger. One of the best-tested consensus algorithms which have demonstrated robustness and security is Proof-of-Work (PoW) \cite{10.1007/3-540-48071-4_10}. PoW relies on validating a proposed block of new transactions to be added to the blockchain by selecting and rewarding a successful ``miner'' who is the first to solve a computational puzzle. This puzzle involves a one-way function, i.e. a function that is easy to compute, and hence easy to verify, but hard to invert. Traditionally the chosen function is the inverse hashing problem, which by its structure makes the parameters of the problem dependent on the current block information, thus making pre-computation infeasible. Additionally, the problem is progress-free, meaning the probability of successfully mining a block at any given instant is independent of prior mining attempts. This means a miner's success probability essentially grows linearly with the time spent, or equivalently work expended, solving the problem. The latter feature ensures that the mining advantage is proportionate to a miner's hashing power.

There are, however, two issues that threaten to compromise the continued usage of PoW consensus in a scalable manner. The first is energy consumption. Problems like inverse hashing admit fast processing, now at speeds of tens of THash/s, by application-specific integrated circuits (ASICs). Unfortunately, the tremendous speed of these devices comes at the cost of large power consumption, and as the hashing power of the network grows, so does the energy cost per transaction. The reason is for asset-based cryptocurrencies like Bitcoin, as the overall network hashing power grows, the difficulty of the one-way function is increased to maintain a constant transaction speed. Since new bitcoins are introduced through the mining process, a constant transaction speed is desirable to maintain stability and avoid inflationary pressures. As of May 2023, a single Bitcoin transaction had the equivalent energy consumption of an average U.S. household over 19.1 days \href{https://digiconomist.net/bitcoin-energy-consumption}{(Digiconomist)}.




The energy consumption of PoW blockchains can be seen as wasteful and unnecessary, given that there are alternative consensus mechanisms, such as Proof-of-Stake (PoS), that require significantly less energy to operate. However, PoS has some other liabilities such as the plutocratic feature of mining power being dependent on the number of coins held by a miner, and vulnerability to so-called ``long-range'' and ``nothing at stake'' attacks \cite{LRA}. 
As a result, there have been growing calls for the development of more sustainable and environmentally friendly PoW blockchain technologies.

The second issue is that PoW assumes only classical computers are available as mining resources. Quantum computing technology, while only at the prototype stage now, is rapidly developing. Quantum computers running Grover's search algorithm \cite{Grover}, can achieve a quadratic speedup in solving unstructured problems like inverting one-way functions. This means if they were integrated into PoW, the progress-free condition would no longer apply and the probability of solving the problem grows super-linearly with computational time spent \footnote{Specifically the probability to solve in time $t$ grows like \mbox{$p(t)=\sin^2(ct)$}, where \mbox{$c=O(\sqrt{D/H})$}, $H$ is the size of the search domain for the one-way function, and $D$ is the number of satisfying arguments.}. This can distort network dynamics in a variety of ways. One example \cite{Sattath2020} is the case of an isolated quantum miner competing against classical miners. Normally, after a block has been successfully won but not verified by the entire network a miner who receives this block would cease working on it and start mining a new block on top of it. Instead, a quantum miner in the middle of a Grover search can choose to measure their register mid-computation, and if they succeed, broadcast the answer to the network. Because of imperfect connectivity, their solution might be accepted by a majority of the nodes instead of the first solver. Another scenario is a network with all quantum miners. Because of partial progress, the miners have a chance to win early and a mixed Nash equilibrium strategy results which favours a probabilistic distribution of times to measure~\cite{Lee2018StrategiesFQ}. This can introduce substantial fluctuations in the time to verify transactions. Workarounds can be found, such as requiring the miners to commit to a time to mine before they begin~\cite{Sattath2020} or using random beacons that interrupt the search progress of quantum computers by periodically announcing new puzzles to be solved~\cite{cryptoeprint:2022/1423}. However, time stamps can be spoofed and as quantum computers speed up and are parallelized, the frequency of interrupting beacons will need to increase to avoid distortions in the consensus dynamics. A future-proofed consensus algorithm should take quantum processing into account as a core resource. 


We propose a new PoW consensus protocol based on boson sampling. Boson-sampling was originally developed to demonstrate \emph{quantum supremacy}, owing to its reduced resource requirements compared to the other quantum algorithms \cite{bib:aa}. Boson-samplers are specialized photonic devices that are restricted in the sense that they are neither capable of universal quantum computing nor error correctable, through proposals have been made to find practical applications in chemistry, many-body physics, and computer science \cite{Lau2022}. We formulate a practical application of a boson-sampling variant called coarse-grained boson-sampling (CGBS) \cite{PhysRevA.94.012315, Nikolopoulos2019}. This scheme involves the equal-size grouping of the output statistics of a boson sampler into a fixed number of bins according to some announced binning tactic. 


The advantage provided by binning the output probability distribution is the polynomial number of samples required to verify a fundamental property of the distribution as opposed to the exponential samples required when no binning is performed. While boson-samplers are not arbitrarily scalable owing to lack of error correction, we argue nonetheless the speedup provided is dramatic enough to warrant their use for PoW consensus.

Photonic-based blockchain has been investigated before. Optical PoW \cite{OPoW2020} uses HeavyHash, a slight modification of the Bitcoin protocol, where a photonic mesh-based matrix-vector product is inserted in the middle of mining. This has already been integrated into the cryptocurrencies optical Bitcoin and Kaspa. Recently, a more time and energy-efficient variant named LightHash has been tested on networks of up to $4$ photons \cite{LightHash}. Both of these protocols use passive linear optics networks acting upon coherent state inputs which implement matrix multiplication on the vector of coherent amplitudes. It is conjectured that the photonic implementation of this matrix multiplication can achieve an order of magnitude speedup over traditional CPU hardware. They exploit the classical speedup associated with a photonic implementation of this operation and do not exploit any quantum advantage. While that method uses a multi-mode interferometer similar to what we describe in this work, it does not use intrinsically quantum states of light and in fact, is a different form of classical computing using light. In contrast, our boson sampling method uses quantum resources with processes that become exponentially harder, in the number of photons, to simulate with classical hardware whether photonic or not.



\section{Background}

\subsection{Boson-sampling} \label{BSfundamentals}

Boson-sampling \cite{bib:aa, bib:bs_intro} is the problem of sampling multi-mode photo-statistics at the output of a randomised optical interferometer. This problem constitutes a noisy intermediate scale quantum (NISQ) \cite{Preskill2018quantumcomputingin} protocol, naturally suited to photonic implementation. Like other NISQ protocols, boson sampling is not believed to be universal for quantum computation, nor does it rely on error correction, thereby limiting scalability. Nonetheless, it has been shown\footnote{Under reasonable complexity-theoretic assumptions.} to be a classically inefficient yet quantum mechanically efficient protocol, making it suitable for demonstrating \emph{quantum supremacy}, which is now believed to have been achieved \cite{bib:china_bs, bib:xanadu_bs}.

Unlike \emph{decision problems}, which provide a definitive answer to a question, boson-sampling is a \emph{sampling problem} where the goal is to take measurement samples from the large superposition state exiting the device.
Since boson-sampling is not an \textbf{NP} problem \cite{bib:comp_complex}, the full problem cannot be efficiently verified by classical or quantum computers. Indeed, even another identical boson sampler cannot be used for verification since results are probabilistic and in general unique, ruling out a direct comparison of results as a means of verification. Nonetheless, restricted versions of the problem such as coarse-grained boson sampling, described below, can be used for verification.

\subsubsection{Fundamentals}

\begin{figure}[t!]
    \centering
    \includegraphics[width=\columnwidth]{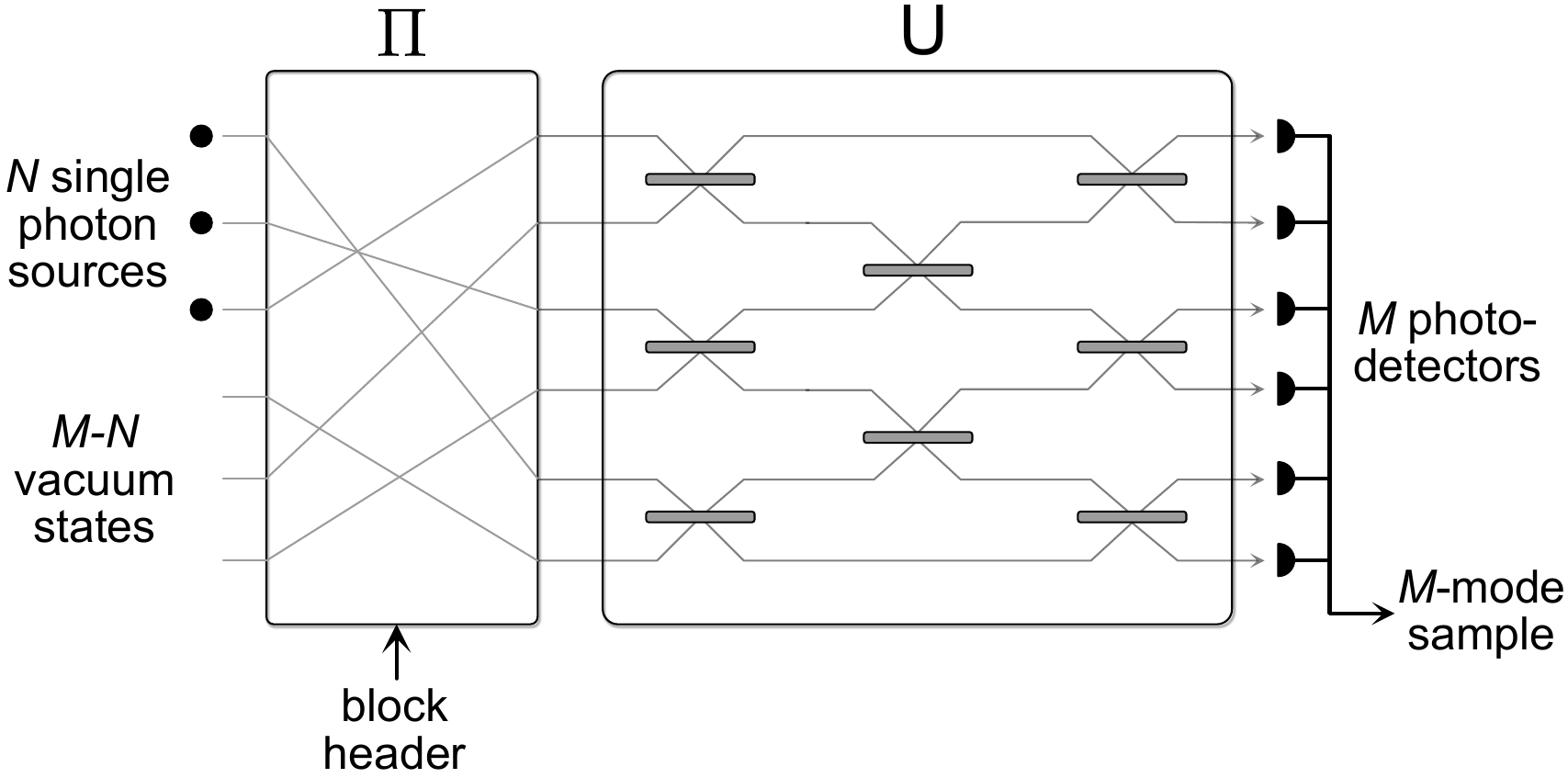}
    \caption{Illustration of the use of a boson-sampling device for blockchain consensus. Initially, $N$ photons are incident in the first $N$ modes, with the remaining \mbox{$M-N$} modes in the vacuum state. The modes then undergo a permutation $\Pi$ dependent on the block header information, which in practice would be accomplished by simply permuting the locations of the single-photon inputs. The photons then pass through an interferometer circuit of depth $M$ described by unitary $U$. Finally, the photons are detected at the $M$ output ports providing a measurement record of the sample.}
    \label{fig:boson_sampling_permutation}
\end{figure}

The general setup for the boson sampling problem is illustrated in Fig.~\ref{fig:boson_sampling_permutation}. We take $M$ optical modes of which $N$ are initialised with the single-photon state and \mbox{$M-N$} with the vacuum state at the input,
\begin{align}
    \ket{S} &= \ket{1}^{\otimes N}\otimes\ket{0}^{\otimes M-N} \nonumber\\
    &= \hat{a}^\dag_1\dots\hat{a}^\dag_N\ket{0}^{\otimes M},
\end{align}
where $\hat{a}^\dag_i$ is the photonic creation operator on the $i$th mode.
Choosing \mbox{$M\geq O(N^2)$} ensures that with a high likelihood the output state remains in the anti-bunched regime whereby modes are occupied by at most one photon. Hence, such samples may be represented as $m$-bit binary strings.

The input state is evolved via passive linear optics comprising beamsplitters and phase-shifters, implementing the Heisenberg transformation on the photonic creation operators,
\begin{align} \label{eq:lo_transform}
    \hat{U} \hat{a}^\dag_i \hat{U}^\dag \to \sum_{j=1}^M U_{i,j} \hat{a}^\dag_j,
\end{align}
where $U$ is the \mbox{$M\times M$} unitary matrix representing the multi-mode linear optics transformation \footnote{To be distinguished from the operator (with a hat) $\hat{U}$ that is an exponential of a bilinear form of creation and annihilation operators.}. That is, each input photonic creation operator is mapped to a linear combination of creation operators over the output modes.

The linear optics transformation $U$ is chosen uniformly at random from the Haar measure, which is essential to the underlying theoretical complexity proof. It was shown by \cite{bib:reck} that any \mbox{$M\times M$} linear optics transformation of the form shown in Eq.~\ref{eq:lo_transform} can be decomposed into a network of at most $O(M^2)$ beamsplitters and phase-shifters, ensuring that efficient physical implementation is always possible. As presented in Fig.~\ref{fig:boson_sampling_permutation}, the number of detectors equals the number of modes $M$. In practice, the number of detectors can be reduced by exploiting multiplexing in other degrees of freedom, such as the temporal degree of freedom. For example, in the architecture presented in Ref.~\cite{PhysRevLett.113.120501}, where modes are encoded temporally, a single time-resolved detector is sufficient for detecting and distinguishing between all modes. 

The output state takes the general form,
\begin{align} \label{eq:bs_output_operator}
    \ket{\psi}_\mathrm{out} &= \left[ \prod_{i=1}^N \sum_{j=1}^M U_{i,j} \hat{a}^\dag_j \right] \ket{0}^{\otimes M} \\
    &= \sum_{k=1}^{|Y|}\alpha_k \ket{Y_k},\nonumber
\end{align}
where \mbox{$\ket{Y_k}=\ket{y_1^{(k)},\dots,y_M^{(k)}}$} denotes the occupation number representation of the $k$th term in the superposition with $y_i^{(k)}$ photons in the $i$th mode, and $\alpha_k$ is the respective quantum amplitude, where for normalisation,
\begin{align}
    \sum_{k=1}^{|Y|}|\alpha_k|^2=1.
\end{align}

The number of terms in the superposition is given by,
\begin{align}
    |Y|=\binom{M+N-1}{N}, \label{eq:dimension}
\end{align}
which grows super-exponentially with $M$ in the \mbox{$M\geq O(N^2)$} regime. Since we are restricted to measuring a number of samples polynomial in $N$ from an exponentially large sample space, we are effectively guaranteed to never measure the same output configuration multiple times. Hence, the boson-sampling problem is \emph{not} to reconstruct the full photon-number distribution given in Eq.~\ref{eq:bs_output_operator}, but rather to incompletely sample from it.

In the lossless case, the total photon number is conserved. Hence,
\begin{align}
\sum_{i=1}^M x_i = \sum_{i=1}^M y_i^{(k)} = N \,\,\forall\,\, X,Y,k,
\end{align}
where \mbox{$\ket{X}=\ket{x_1,\dots,x_M}$} represents the occupation number representation of the input state.

The amplitudes in the output superposition state are given by,
\begin{align}
    \alpha_k = \bra{Y_k} \hat{U} \ket{X} = \frac{{{\mathrm{Per}}({U_{X,{Y_k}}})}}{{\sqrt {\prod\nolimits_{i = 1}^M {{x_i}!{y_i^{(k)}}!}}}},
\end{align}
where $\mathrm{Per}(\cdot)$ denotes the matrix permanent, and $U_{X,Y}$ is an \mbox{$N\times N$} sub-matrix of $U$ composed by taking $x_i$ copies of each row and $y_i^{(k)}$ copies of each column of $U$. The permanent arises from the combinatorics associated with the multinomial expansion of Eq.~\ref{eq:bs_output_operator}, which effectively sums the amplitudes over all possible paths input photons $X$ may take to arrive at a given output configuration $Y_k$.

The probability of measuring a given output configuration $Y_k$ is simply,
\begin{align}
    {\rm Pr}(Y_k) = |\alpha_k|^2.
\end{align}
In lossy systems with uniform per-photon loss $\eta$, all probabilities acquire an additional factor of $\eta^N$ upon post-selecting on a total of $N$ measured photons,
\begin{align}
    {\rm Pr}(Y_k) = \eta^N |\alpha_k|^2.
\end{align}
The overall success probability of the device is similarly,
\begin{align}
    \mathrm{Pr}_\mathrm{success}=\eta^N.
\end{align}

Calculating matrix permanents is \textbf{\#P}-hard in general, a complexity class even harder than \textbf{NP}-hard\footnote{\textbf{\#P} is the class of counting problem equivalents of \textbf{NP} decision problems. While the class of \textbf{NP} problems can be defined as finding a satisfying input to a boolean circuit yielding a given output, \textbf{\#P} enumerates \emph{all} satisfying inputs.}, from which the classical hardness of this sampling problem arises. It should however be noted that boson-sampling does not let us efficiently \emph{calculate} matrix permanents as this would require knowing individual amplitudes $\alpha_k$. The $\alpha_k$ amplitudes cannot be efficiently measured since we are only able to sample a polynomial subset of an exponentially large sample space, effectively imposing binary accuracy as any output configuration is unlikely to be measured more than once.

\subsubsection{Mode-binned boson-sampling} \label{sec.mode-bin}

Consider an $N$-photon, $M$-mode boson-sampling experiment where the output modes are arranged in $d^{(\mathtt{mb})}$ bins labelled \mbox{$\mathtt{bin}^{(\mathtt{mb})}_1, \mathtt{bin}^{(\mathtt{mb})}_2, \ldots, \mathtt{bin}^{(\mathtt{mb})}_{d^{(\mathtt{mb})}}$}. Given a linear optical unitary $\hat{U}$ on $M$ modes, let $P({\bf n})$ be the probability of measuring the multi-photon binned number output described by the output vector
\begin{align}
    {\bf n}=(n_1,n_2,\ldots, n_{d^{(\mathtt{mb})}}),
\label{ouputbinvec}
\end{align}
with $n_i$ photons in $\mathtt{bin}_i$. It was shown in \cite{SNAC22} that this distribution can be expressed as the discrete Fourier transform over the characteristic function, 
\begin{align}
    P^{({\mathtt{mb}})}({\bf n})=\frac{1}{(N+1)^{d^{(\mathtt{mb})}}}\sum_{{\bf c}\in \mathbb{Z}_{N+1}^{d^{(\mathtt{mb})}}}\chi\left(\frac{2\pi {\bf c}}{N+1}\right)e^{-i \frac{2\pi {\bf c}\cdot {\bf n}}{N+1} },\label{charfunc}
\end{align}
where
\begin{align}
    \chi({\bf s})=\bra{\Psi_{{\rm in}}}\hat{U}^{\dagger}e^{i 2\pi {\bf s}\cdot \hat{{\bf N}}_{d^{(\mathtt{mb})}}}  \hat{U}\ket{\Psi_{{\rm in}}},
\end{align}
and the vector of binned number operators is,
\begin{align}
    \hat{{\bf N}}_{d^{(\mathtt{mb})}}=\left(\sum_{j_1\in {\mathtt{bin}}^{(\mathtt{mb})}_1} \hat{n}_{j_1},\ldots , \sum_{j_{d^{(\mathtt{mb})}}\in {\mathtt{bin}}^{(\mathtt{mb})}_{d^{(\mathtt{mb})}}} \hat{n}_{j_{d^{(\mathtt{mb})}}}\right).\label{eq:binnedN}
\end{align}

The characteristic function can be computed directly as a matrix permanent,
\begin{align}
    \chi({\bf s})={\rm Per}(V_N({\bf s})),
\end{align}
with
\begin{align}
    V({\bf s})=U^{\dagger}D({\bf s})U,
\end{align}
where the diagonal matrix \mbox{$D({\bf s})=\prod_{j=1}^{d^{(\mathtt{mb})}} D^{(j)}(s_j)$} and
\begin{align}
[D^{(j)}(s_j)]_{u,v}=
\left\{\begin{array}{ccc}1\ &{\rm if}&\ u=v\  {\rm and }\ u\not\in {\mathtt{bin}}^{(\mathtt{mb})}_j \\e^{is_j}\ &{\rm if}&\ u=v\  {\rm and }\ u\in {\mathtt{bin}}^{(\mathtt{mb})}_j\\0\ &{\rm if}&\ u\neq v\end{array}.\right.
\end{align}
Here $V_N({\bf s})$ means taking the \mbox{$N\times N$} matrix formed from the $N$ rows and $N$ columns of the \mbox{$M\times M$} matrix $V$ according to the mode location of single-photon inputs in the input vector $\ket{\Psi_{{\rm in}}}$.

By Eq.~\ref{charfunc}, the mode-binned probability distribution can be computed by evaluating 
$\binom{N+d^{(\mathtt{mb})}-1}{N}$ permanents. To exactly compute the permanent of an \mbox{$N\times N$} matrix requires $O(N2^N)$ elementary operations using Ryser's algorithm, but if we only demand a polynomial additive approximation then a cheaper computational method is available. We can use the Gurvits' approximation which allows for classical estimation of the permanent of a complex \mbox{$N\times N$} matrix to within additive error $\delta$ in \mbox{$O(N^2/\delta^2)$} operations. The algorithm works by sampling random binary vectors and computing a Glynn estimator (Appendix \ref{appendix:modebinest}). The number of random samples $m$ needed to approximate $\chi({\bf s})$ to within $\delta$ with probability at least $p$ is
\begin{align}
    m=\frac{2}{\delta^2}\ln(2/(1-p)),
\end{align}
and each Glynn estimator can be computed in $N^2$ elementary steps.
We now introduce the total variation distance between two distributions with support in some domain $D$ defined
\begin{align}
    \mathcal{D}^{({\rm tv})}(P,Q)\equiv\frac{1}{2}\sum_{{\bf x}\in D}|P({\bf x})-Q({\bf x})|.
\end{align}
The motivation for using the total variation distance here and in the remainder of the protocol is the following. Consider the problem of deciding if a data set consisting of identically and independently distributed (iid) random variables should be considered to be drawn from one distribution $P$ or another distribution $Q$. When the prior is unbiased, then the Bayes error $P_e$ of incorrectly classifying the data is given by the simple relation $P_e=\frac{1}{2}(1-\mathcal{D}^{({\rm tv})}(P, Q))$ \cite{NIELSEN201425}.

By choosing 
\begin{align}
    \delta \leq \frac{\beta}{\sqrt{\binom{N+d^{(\mathtt{mb})}-1}{N}}},
\end{align}
an estimate $\widehat{P^{({\mathtt{mb}})}}({\bf n})$ of the mode-binned distribution can be obtained such that \mbox{$\mathcal{D}^{({\rm tv})}(\widehat{P^{({\mathtt{mb}})}},P^{({\mathtt{mb}})})\leq \beta$}. The number of elementary operations to compute this estimate is\footnote{We ignore the cost to compute the $M\times M$ matrices $V({\bf s})$ as this could be pre-computed for all ${\bf s}$ since we assume a fixed unitary $U$ in the protocol to follow.}
\begin{align}
    \frac{2\ln(2/(1-p)) N^{2}\binom{N+d^{(\mathtt{mb})}-1}{N}^2\log(N)}{\beta^{2}}.
\end{align}
For a fixed number of bins $d^{(\mathtt{mb})}$, this provides a classical polynomial time in $N$ approximation to the mode-binned distribution. Regarding the number of quantum samples needed, it has been shown \cite{VV17} that if one has the means to draw samples from a distribution $Q$, the number of samples $N_{\rm tot}$ needed to distinguish $Q$ from another distribution $P$ is 
\begin{align}
    \frac{c\sqrt{|D|}}{\mathcal{D}^{({\rm tv})}(Q,P)^2}.
    \label{eq:num_tot}
\end{align}
Here, choosing the constant $c=2^{16}$ assures that the test succeeds with probability at least $3/4$. For the mode-binned boson-sampling distribution, we can choose $Q$ to be the distribution from which nodes are sampling from $P_{\mathtt{BS}}^{(\mathtt{mb})}({\bf n})$, and $P$ to be the estimate of the true distribution $\widehat{P^{({\mathtt{mb}})}}({\bf n})$. The dimension $|D|$ will be the total number of ways $N$ photons can be put in $d^{(\mathtt{mb})}$ bins. This is given by 
\begin{align}
    |D|=\binom{N+d^{(\mathtt{mb})}-1}{N}.
\end{align}

We want to guarantee that the following cases are rejected
\begin{align} \label{eq:reject}
    \mathcal{D}^{({\rm tv})}(P^{({\mathtt{mb}})}({\bf n}),P_{\mathtt{BS}}^{(\mathtt{mb})}({\bf n})) \geq \beta.
\end{align}

Since the total variation distance is a distance metric, we can write
\begin{align}
    &\mathcal{D}^{({\rm tv})}(P^{({\mathtt{mb}})}({\bf n}),P_{\mathtt{BS}}^{(\mathtt{mb})}({\bf n})) \\ \nonumber
    &\quad \geq \mathcal{D}^{({\rm tv})}(P_{\mathtt{BS}}^{(\mathtt{mb})}({\bf n}),\widehat{P^{({\mathtt{mb}})}}({\bf n})) - \mathcal{D}^{({\rm tv})}(P^{({\mathtt{mb}})}({\bf n}),\widehat{P^{({\mathtt{mb}})}}({\bf n}))  \\ \nonumber
    &\quad \geq \mathcal{D}^{({\rm tv})}(P_{\mathtt{BS}}^{(\mathtt{mb})}({\bf n}),\widehat{P^{({\mathtt{mb}})}}({\bf n})) - \beta,
\end{align}
where we have used the fact that $\mathcal{D}^{({\rm tv})}(P^{({\mathtt{mb}})},\widehat{P^{({\mathtt{mb}})}}({\bf n})) \leq \beta$. So in order to reject cases in Eq.~\ref{eq:reject}, the following has to be true
\begin{align}
  \mathcal{D}^{({\rm tv})}(\widehat{P^{({\mathtt{mb}})}}({\bf n}),P_{\mathtt{BS}}^{(\mathtt{mb})}({\bf n})) \geq 2\beta  
\end{align}

The number of samples needed to distinguish the estimate $P^{(\mathtt{mb})}$ from $P_{\mathtt{BS}}$ that is more than $2\beta$ in total variation distance away is
\begin{align}
    N^{(\mathtt{mb})}_{{\rm tot}}=2^{14}\frac{\sqrt{\binom{N+d^{(\mathtt{mb})}-1}{N}}}{\beta^2}.
    \label{eq:Ntot}
\end{align}

\begin{figure}[!htp]
    \includegraphics[width=.8\linewidth]{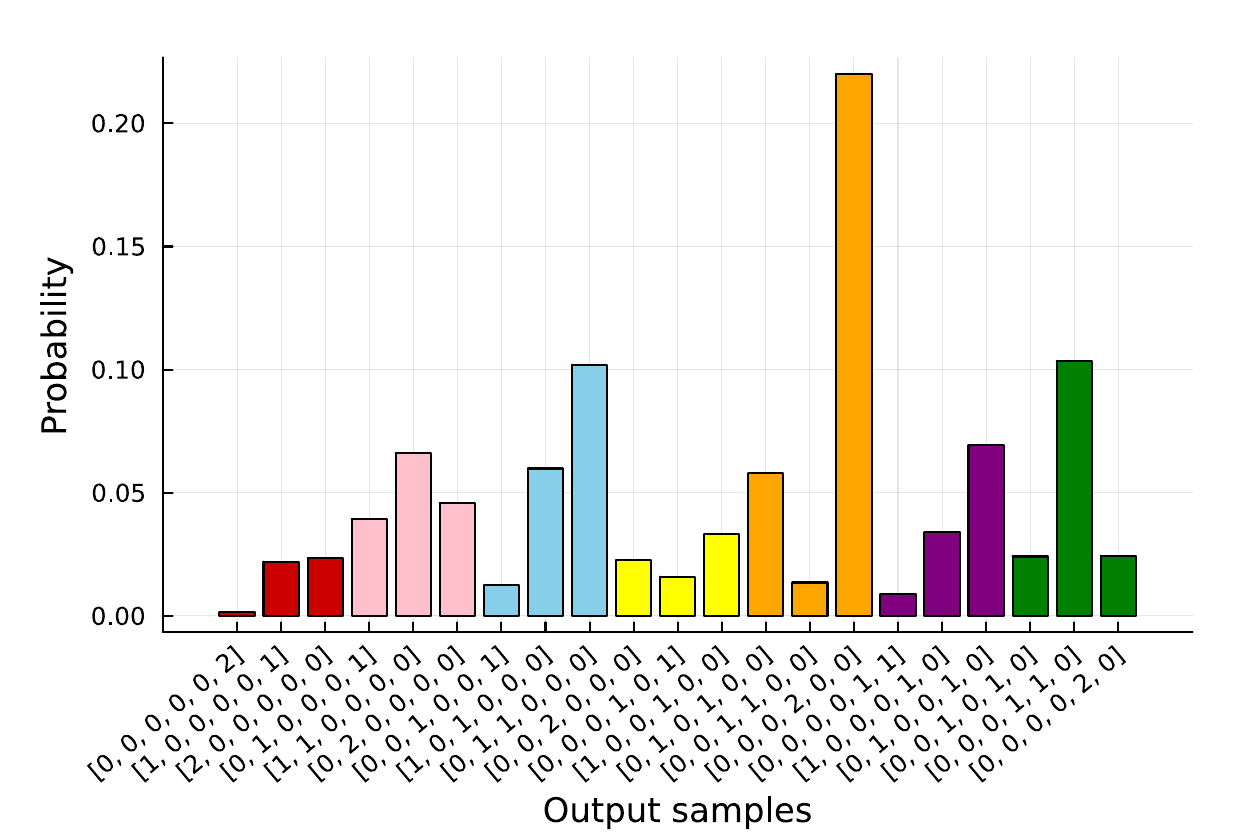} \\
    \includegraphics[width=.8\linewidth]{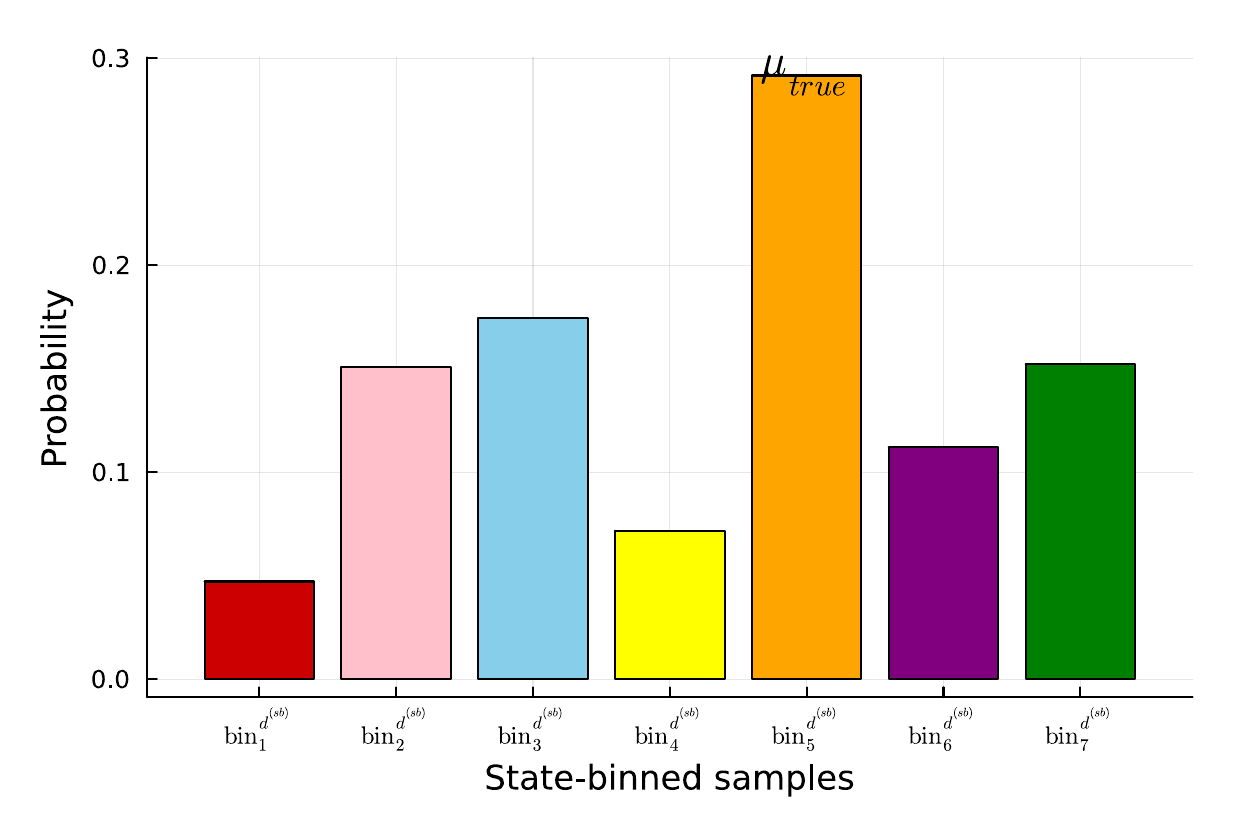} \\
    \includegraphics[width=.8\linewidth]{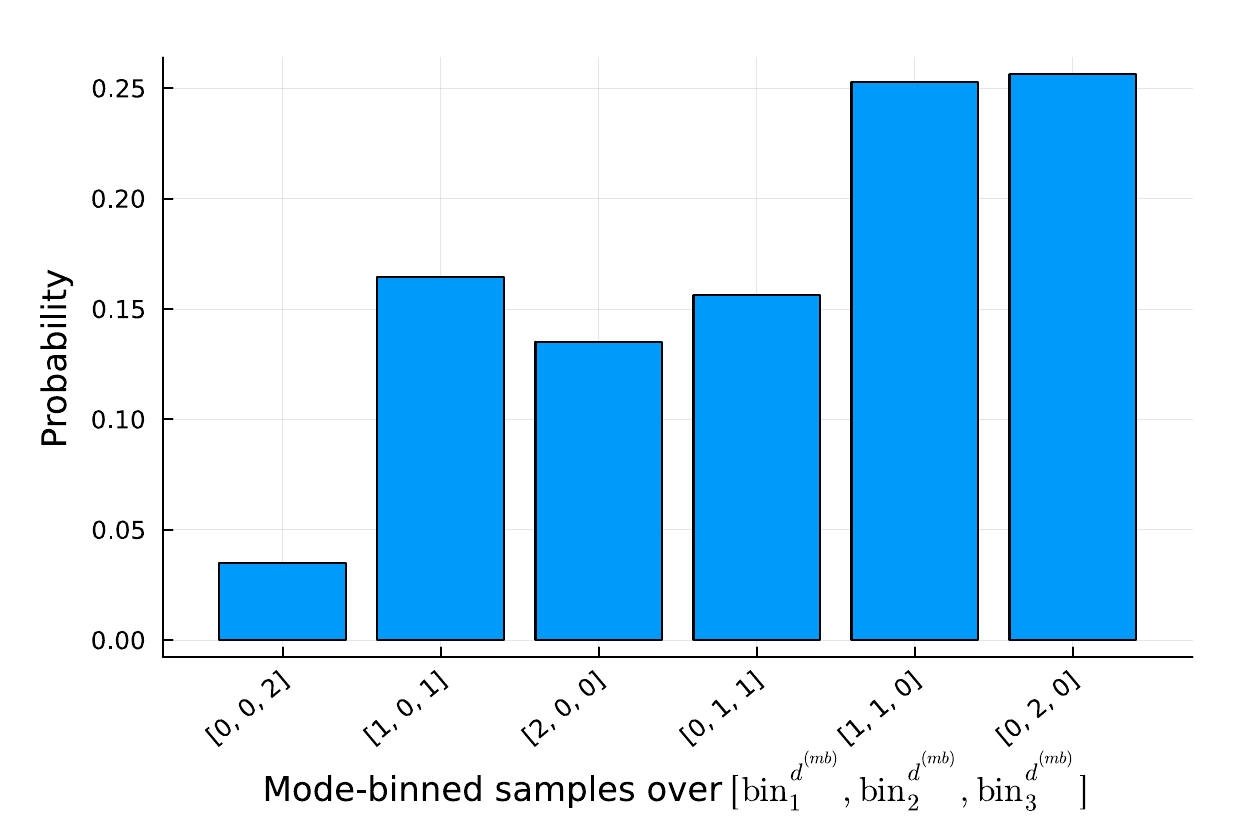}
    \caption{Plots showing the output probability distribution of a Haar random boson-sampling device with two photons in six modes, i.e. $N=2$ and $M=6$, for which a total of $\binom{6+2-1}{2}$, i.e. 21 output photon configurations are possible. (a) BS distribution without any binning. The x-axis shows the different ways in which two photons can exit the six modes of the boson sampler and the corresponding probabilities of these configurations. (b) State-binned distribution of the same experiment where the 21-dimensional output Hilbert space is binned into $d^{(\mathtt{sb})}=7$ bins, each bin containing three configurations chosen by the colour code as visible in both (a) and (b). Note that $\mathtt{bin}^{d^{(\mathtt{sb})}}_{5}$ has the maximum probability of $\mu_{\mathtt{true}} = 0.29$.  (c) Mode-binned distribution of the same experiment where the modes are grouped into $d^{(\mathtt{mb})} = 3$ bins, each mode bin containing two consecutive bins. A total of $\binom{3+2-1}{2}$, i.e. 6 output photon configurations are possible for this mode-binning. In these figures, we have only plotted the physical configurations where the total photon number is conserved out of the total configuration space.}
    \label{fig:binning}
\end{figure}
In practice the number of samples required is much smaller as shown in Fig.~\ref{fig:num_samples_N}. See Sec.~\ref{sec:Robustness} for more detailed discussions on this.
\subsubsection{State-binned boson-sampling} \label{sec:sbbs}

An alternative to the above procedure where bins are defined by sets of output modes is to bin according to sets of multimode Fock states. For an $N$-photon input state in an $M$-mode unitary $U$, the number of possible output configurations is given by $|Y|$ as defined in Eq.~\ref{eq:dimension}. State-binned boson sampling then concerns the binning of this $|Y|$ dimensional Hilbert space into $d^{(\mathtt{sb})}$ bins. 

For a given boson-sampling experiment, the output samples are essentially the $\vert Y_{k} \rangle$ configuration vectors as defined in Eq.~\ref{eq:bs_output_operator}, where $1 \leq k \leq \binom{N+M-1}{N} $. However, the state-binned samples into $d^{(\mathtt{sb})}$ bins, on the same Boson-sampling experiment are given by the $\vert bin_{l}^{(\mathtt{sb})} \rangle$ configuration vectors, where 
\begin{align}
    \vert bin_{l}^{(\mathtt{sb})} \rangle = \bigcup_{j} \vert Y_{j} \rangle,
\end{align}
and the union over $j$ can be chosen according to any agreed-upon strategy such that $1 \leq l \leq d^{(\mathtt{sb})}$. In this paper, we consider the case where all bins contain an equal number of configuration vectors. 

Given any binning strategy, the bin with the maximum probability is defined as $bin_{true}^{d^{(\mathtt{sb})}}$, and the corresponding peak bin probability (PBP) is defined as $\mu_{\mathtt{true}}$. If the complete output bin probability distribution is unknown, the PBP $\mu_{\mathtt{net}}$ of the incomplete probability distribution serves as an estimate of $\mu_{\mathtt{true}}$. That is, assuming that the honest nodes on the blockchain network provide enough samples for the same boson-sampling experiment, the PBP $\mu_{\mathtt{net}}$ will be a close approximation to the PBP $\mu_{{\rm true}}$ of the binned boson-sampling problem. 


Specifically, we wish to ensure that
\begin{align}
    {\rm Pr}[\mu_{\mathtt{net}}-\epsilon/2<\mu_{{\rm true}}<\mu_{\mathtt{net}}+\epsilon/2]>1-\gamma,
\end{align}
for some accuracy $\epsilon<1/d^{(\mathtt{sb})}\ll 1$ where $\gamma\ll 1 $ determines the $100(1-\gamma)\%$ confidence interval for $\mu_{{\rm true}}$. It was shown in Ref.~\cite{Nikolopoulos2019} that this can be achieved for perfect boson sampling using a sample size of at least
\begin{align}
    N^{(\mathtt{sb})}_{{\rm tot}}=\frac{12 d^{(\mathtt{sb})}}{\epsilon^2}\ln(2\gamma^{-1}).
\end{align}
Using a bootstrap technique obtained by resampling provided samples from the boson-sampling distribution, it is shown \cite{Nikolopoulos2019} that the required accuracy can be obtained when
$2d^{(\mathtt{sb})}\epsilon^{0.8}\lesssim 0.1$,
in which case, if we demand a low uncertainty $\gamma=5\times 10^{-4}$, the number of required samples is
\begin{align}
    N^{(\mathtt{sb})}_{\rm tot}= 1.8\times 10^5 {d^{(\mathtt{sb})}}^{7/2}.
\end{align}

\section{Results}

\begin{figure*}[!ht]
    \centering
    \includegraphics[width=1.0\textwidth]{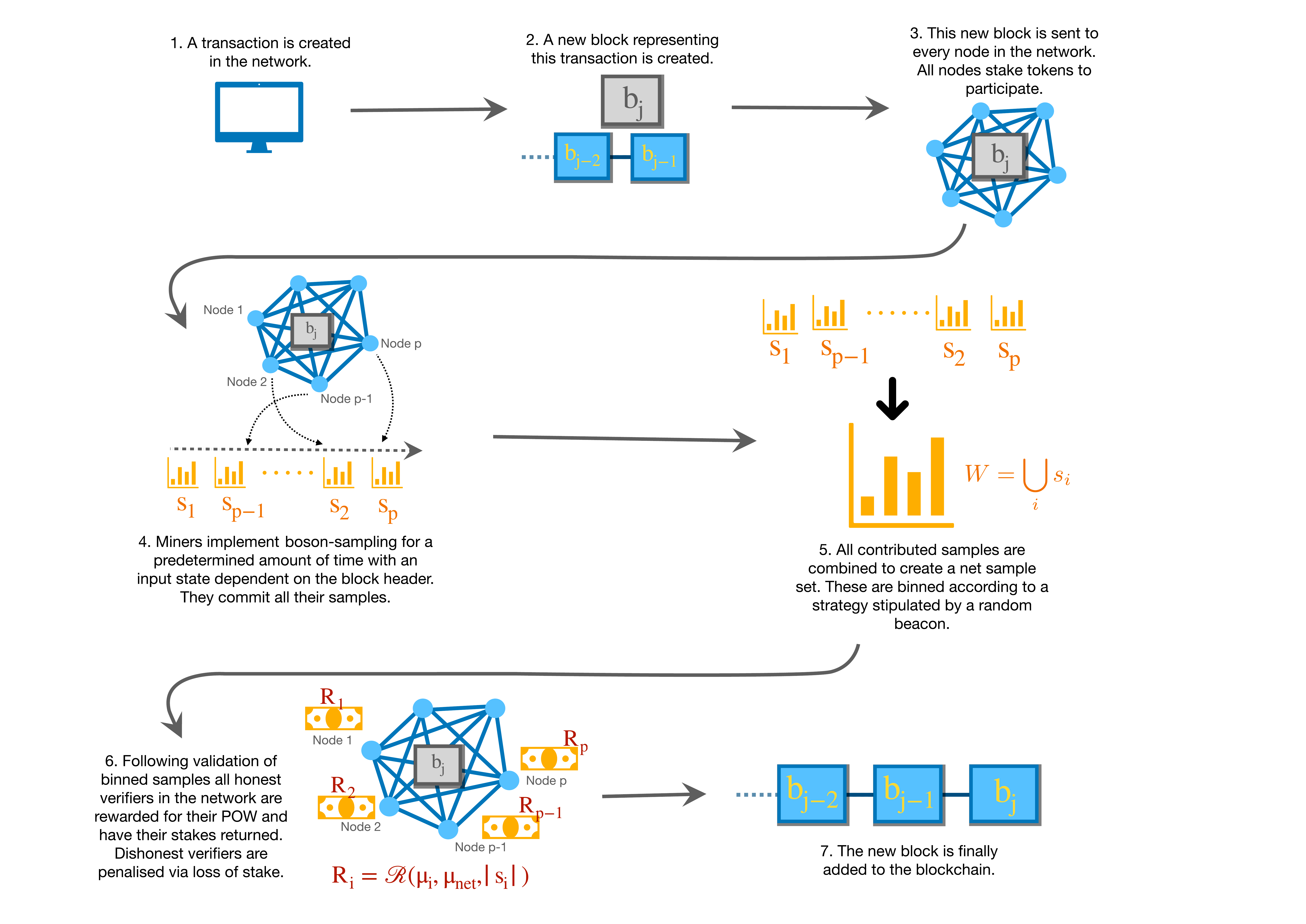}
    \caption{Blockchain architecture with the inclusion of boson-sampling routine.}
    \label{fig:consensus}
\end{figure*}

We utilize two types of binning for PoW consensus, one for validation to catch cheaters, and one to reward miners. The former can be estimated with classical computers efficiently, while the latter does not have a known classical computation though it does have an efficient quantum estimation. Upon successful mining of a block, the output of both binning distributions will be added to the blockchain, meaning one part can be verified efficiently by classical computers while another part cannot. This will incentivize nodes using boson-sampling devices to verify prior blocks in the blockchain. The protocol is illustrated in Fig.~\ref{fig:consensus} and a detailed description is provided in Sec.~\ref{sec:Methods}. See Table~\ref{tab:grouped-rows} in the supporting information text for a description of the various parameters. An alternate approach based on Gaussian boson sampling is described in Sec.~\ref{Sec:GBS}.

\subsection{Robustness}
\label{sec:Robustness}

The key to making this protocol work is that the miners don't have enough information ahead of time about the problem to be solved to be able to pre-compute it but their samples can be validated after they have been committed. The blockchain is tamper-proof because any attempt to alter a transaction in a verified block of the chain will alter that block header and hence the input permutation $\Pi$ that determines the boson-sampling problem and the output record ${\tt Rec}$. One could also use a protocol where the unitary $U$ depends on the block header but it is easier to change the locations of input state photons than to reconfigure the interferometer circuit itself. The number of input states using $N$ single photons in $M$ modes is $\binom{M}{N}$ making precomputation infeasible. 

The record ${\tt Rec}(b_j)$ can be verified since the output distribution $P^{(\mathtt{mb})}$ can be checked in polynomial time (in the number of bins $d^{(\mathtt{mb})}$ and $N$) on a classical computer. The peak probability $\mu_{\rm net}$ can be checked in polynomial time (in the number of bins $d^{(\mathtt{sb})}$) on a quantum boson-sampler. The fact that the miners don't know the mode binning ahead of time, of which there are $M!/(M/d^{(\mathtt{mb})})!^{d^{(\mathtt{mb})}}$
possibilities means that even after the problem is specified, there is no advantage in using even classical supercomputers to estimate $P^{(\mathtt{mb})}$.   
Even if the probability to randomly guess the correct mode binned distribution were non-negligible, for example, because of a choice to use a small $d^{(\mathtt{mb})}$ and large $\beta$ to speed up the validation time, provided it is smaller than $p^{\mathtt{cheat}}$, the protocol is robust. The reason is, as established in Appendix~\ref{payoff}, cheaters will be disincentivized since failure to pass the test incurs a penalty of lost staked tokens. Similarly, not knowing the state-binning means that they have no potential advantage in the payout.

The mining time is
\begin{align}
    T_{\rm mine} = \frac{\max\{N_{tot}^{(\mathtt{mb})},N_{tot}^{(\mathtt{sb})}\}}{R_q},
\end{align}
where $R_q$ is based on publicly available knowledge of the boson sampling repetition rate at the time of the genesis block. This choice for mining time is made to ensure that honest miners with boson samplers will have produced enough samples to pass the validation test and even if there is only one honest node, that node will have produced enough samples to earn a reward. The repetition rate will of course increase with improvements in quantum technology but that can be accommodated by varying the other parameters in the problem such as photon number, bin numbers, and prescribed accuracy, in order to maintain a stable block mining rate. For $N=25$, $d^{(\mathtt{mb})}=3$, and $\beta=0.1$, and assuming the boson sampling specifications in the caption of Fig.~\ref{fig:analysis2}, the minimum mining time would be $81.6$s. The validation test sets a lower limit on the time to execute the entire protocol and add a new block. The classical computation involved during the validation step, while tractable, can be a long computation even for moderate-sized bin numbers $d^{(\mathtt{mb})}$ and photon numbers. However, it is a problem amenable to distributed computation since the problem specification is known to all miners.

The purpose of the state-binning step is twofold. It provides an independent way to tune the reward structure and hence moderate participation in the protocol. Second, it incentivizes nodes to have a quantum boson-sampling device in order to verify older blocks in the blockchain since there is no known efficient classical simulation of the state-binned distribution whereas there is for the counterpart mode-binned distribution under the assumption of a constant number of bins.

In general, the protocol is immune to the most common noise sources in the photonic implementation of boson sampling. Photon loss would not corrupt a node's sample set since only photon number-conserving samples would be committed to the network. The main consequence of loss in an individual node's boson sampler would be to reduce the sampling rate of that node and motivate them to improve their setup. Similarly, unital errors in individual setups only increase the variation distance between the target and erroneous distributions linearly in the number of photons and operator norm distance between the unitaries~\cite{PhysRevA.92.062326}, which is independent of the mode numbers. In a scenario where individual nodes implement unitaries that are $\Delta$ close to the target unitary in the operator norm, the net set of committed samples appears to be a result of a stochastic unitary also $\Delta$ close to the target unitary on average, giving a good approximation of the target probability distribution. Furthermore, in the mode-binning subroutine, a suitable $\beta$ can be fixed to allow for the protocol to be run with partial photon distinguishability but still retaining the computational hardness of boson sampling~\cite{PhysRevResearch.4.023013}. Since sampling from partially distinguishable photons can be expressed in terms of the interference of fewer completely indistinguishable photons~\cite{PhysRevLett.120.220502}, for large enough photon indistinguishability, the sampling distribution is still hard to approximate classically. Specifically, the complexity of the classical algorithm in \cite{PhysRevLett.120.220502} scales as $N^{2k} 2^k k$ for the $k^{\mathrm{th}}$ order approximation of the permanents, which would not be efficient since a value of $k$ close to $N$ would be required for the level of indistinguishability assumed in Sec.~\ref{SampleRates} and already achieved in experiments \cite{deng2023gaussian}. In Fig.~\ref{fig:tvd_distinguibality}, we show how the TVD varies with the indistinguishability parameter defined in Ref \cite{PhysRevLett.120.220502}. As an example (Fig. 4 in Ref \cite{PhysRevLett.120.220502} ) for $N=25$, an accuracy of $0.01$, and indistinguishability of about 0.85 will make the algorithm exponential in its time complexity. If we focus on the same level of indistinguishability in the inset plot in Fig.~\ref{fig:tvd_distinguibality} the TVD is roughly $0.022$. A suitable $\beta$, as discussed above could be chosen such that we accept samples that are closer than this. This is also consistent with the results we present later regarding sensitivity to binning strategy and other classical spoofing methods.

\begin{figure}[!htp]
    \subfloat[]{
        \includegraphics[width=\linewidth]{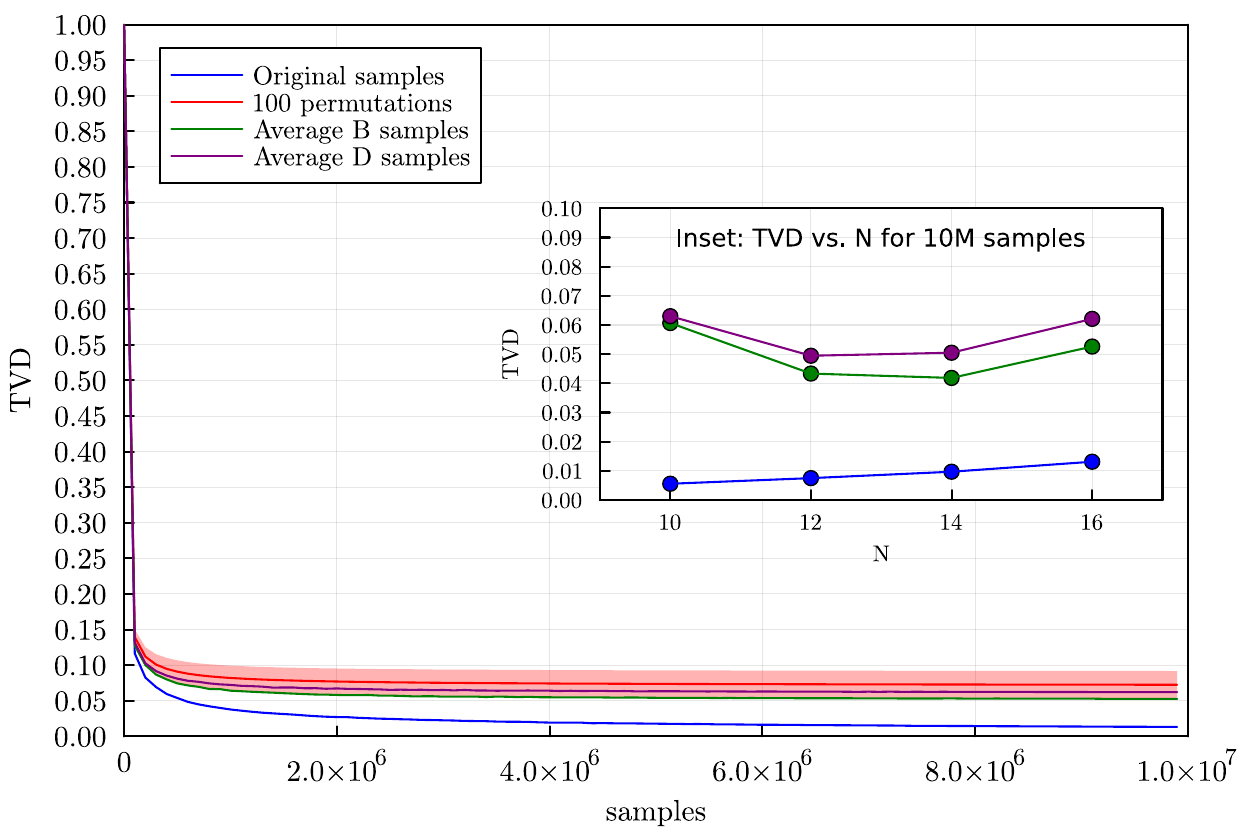}
        \label{fig:plot_inset}
    }
    \hfill
    \subfloat[]{
        \includegraphics[width=\linewidth]{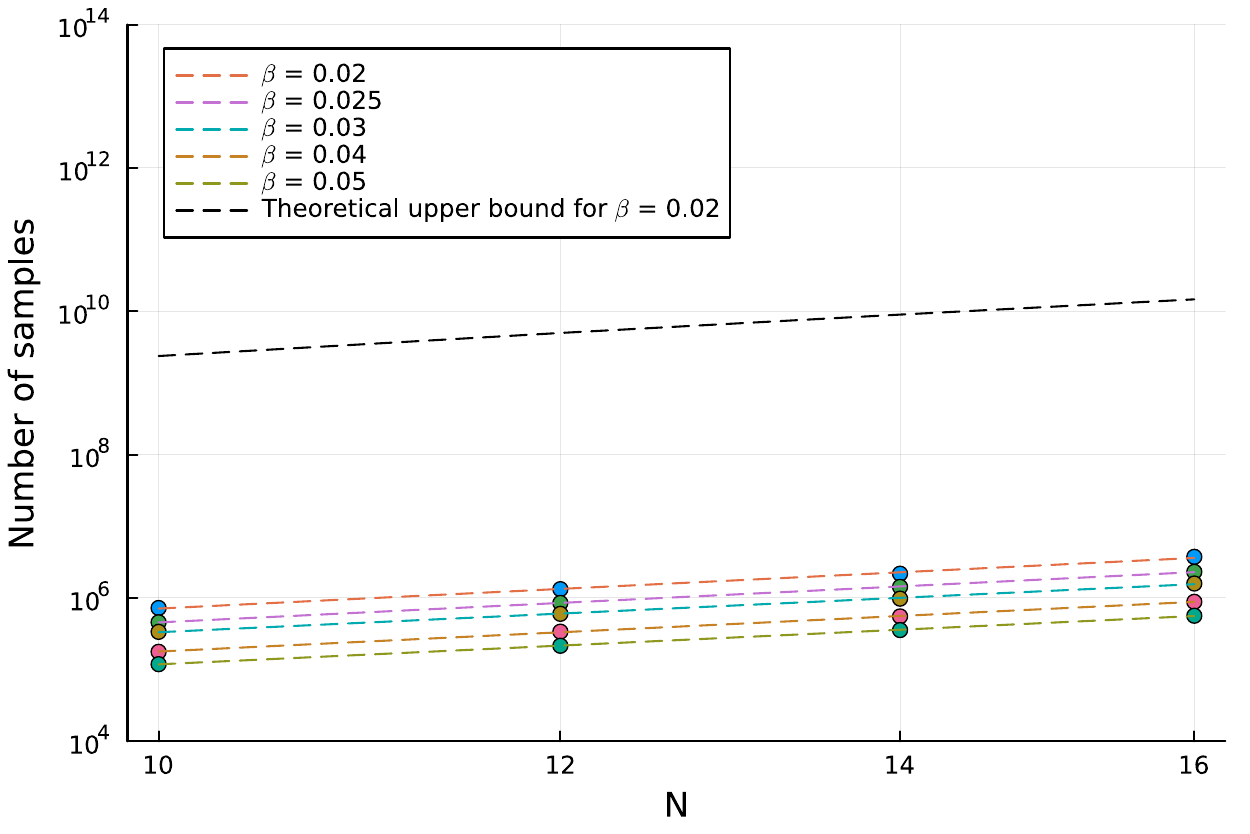}
        \label{fig:num_samples_N}
    }
     
    \caption{Performance of the protocol. (a) Dependence of the total variation distance (TVD) between the exact mode-binned distributions Eq.~\ref{charfunc} and distributions obtained from a finite number of samples for $N=16$ photons generated from the following: the Clifford sampler in BosomSampling.jl package \cite{seron2024bosonsampling} (blue line), the average distribution in the distinguishable case as given by Eq.~\ref{eq:distinguishable_dist} (green line), and the average distribution for the indistinguishable case as given in  Eq.~\ref{eq:bosonic_dist} (purple line). The red line represents the average of the TVD between the exact distribution and the mode-binned distributions generated from a hundred random permutations of the output binning scheme. The inset shows the scaling of the TVD with different photon numbers for 10 million samples. (b) The estimated number of samples for which the TVD between the sample distribution and the exact mode-binned distribution becomes less than a fixed value ($\beta$) Vs N. This is plotted for various $\beta$. The black dashed line represents the theoretical bound given by Eq.~\ref{eq:num_tot} for $\beta = 0.02$. All plots assume a 256-mode interferometer with six bins $(M = 256 \text{ and } d^{(mb)} = 6)$.}
    \label{fig:haaravg}
\end{figure}

\begin{figure}
    \centering
    \includegraphics[width=\linewidth]{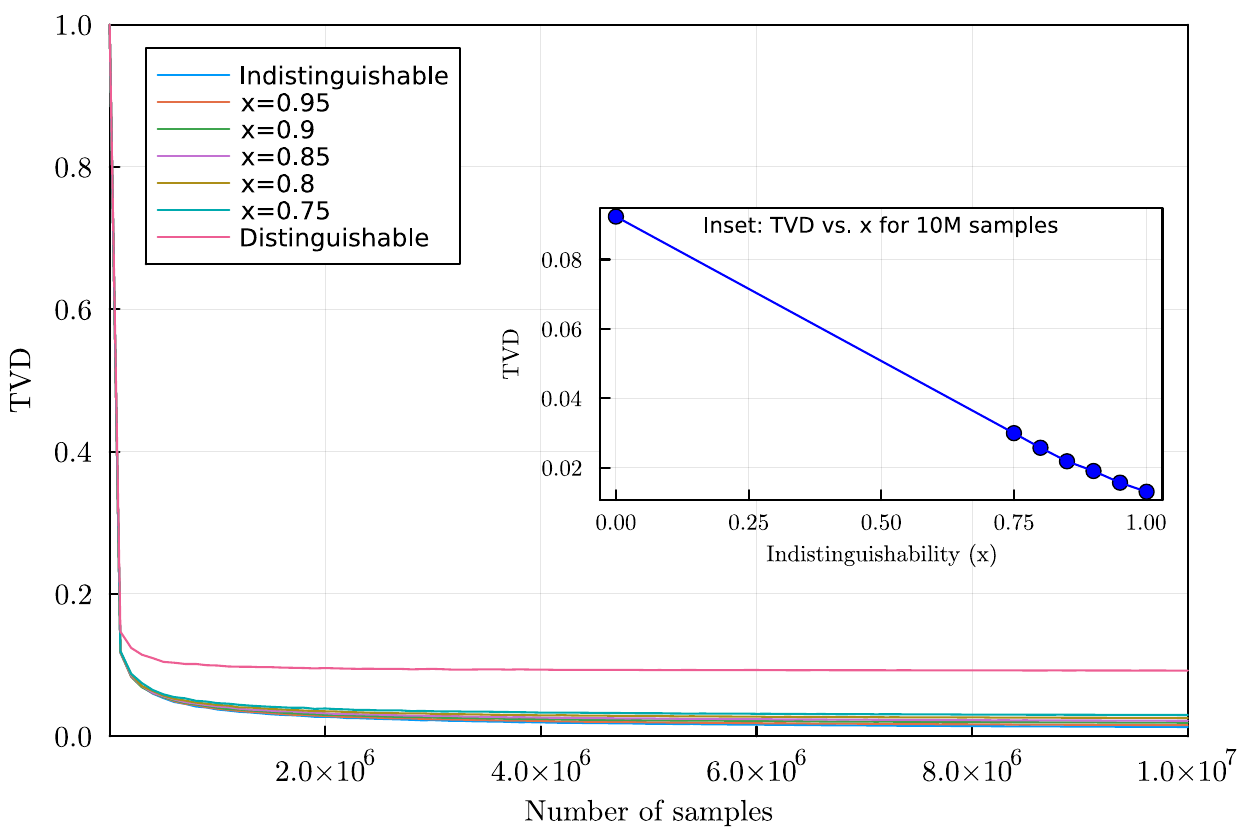}
    \caption{Plot of total variation distance (TVD) between the exact mode-binned distribution and finite sample distribution as a function of the number of samples for different indistinguishability values (x). The indistinguishability parameter ranges from x = 0 (fully distinguishable photons) to x = 1 (fully indistinguishable photons). The inset plot shows how TVD varies with the indistinguishability parameter for a fixed sample size of 10 million. These simulations use $N=16,M=256$, and $d^{(\text{mb})} = 6$. This shows a clear separation of the target from the distribution obtained from fully distinguishable photons and also enables us to select a suitable TVD to reject imperfect samples but accept samples with sufficiently high indistinguishability ($\approx 0.85$) which has been achieved experimentally.}
    \label{fig:tvd_distinguibality}
\end{figure}

In our analysis, we aim for robustness of two kinds: (a) sensitivity to binning strategy and (b) security against Haar-averaged distribution. To analyze the sensitivity of the binning strategy, we collect the samples and calculate the sample distribution for a reference binning strategy. Then we randomly permute the output modes to simulate random binning strategies with the same bin sizes. Note that this numerical simulation can be done on the same sample set as the different binnings are sortings of the mode labels of measured outcomes. We then calculate TVD between the sample distributions of 100 such random permutations and the exact mode-binned distribution with the reference binning strategy. The variance and mean of these TVDs are plotted versus sample size and are shown in Figure~\ref{fig:plot_inset}. It is evident from the plot that the mean of TVDs is well-gapped when compared to the sample distribution with reference binning strategy. This shows that the coarse-grained (marginalized) distribution is sensitive to the binning strategy. This also provides evidence that there exists a $\beta$ such that any classically generated distribution without the knowledge of binning strategy will be at least $\beta$ away in TVD from some reference binned distribution. A naive choice of this $\beta$ would be half of the gap between the sample distribution of reference binning strategy and the mean of permuted ones in Figure~\ref{fig:plot_inset}. This gives us enough confidence to rule out any classical spoofing attack that can well approximate the coarse-grained distribution without prior knowledge of the binning strategy.

Next, we analyze the robustness of our mode-binning validation against potential attacks where a miner submits samples using the mode-binned probability distribution averaged over the Haar-random unitary matrix representing the interferometer \cite{shchesnovich_asymptotic_2017, shchesnovich_quantum_2017,SNAC22}. This distribution can be computed classically by cheaters before the target unitary is even broadcast to the network. These mode-binned distributions are given for the completely distinguishable ($P_D$) and the completely indistinguishable or bosonic case ($P_B$) as follows: 

\begin{align}
    &P_{D}( \textbf{k}) = \frac{n!}{\prod_{z=1}^{d^{(\mathtt{mb})}} k_{z}!} \prod_{z=1}^{d^{(\mathtt{mb})}} q_{z}^{k_{z}}! \label{eq:distinguishable_dist}\\
    &P_{B}( \textbf{k}) = P_{D}(k)\frac{\prod_{z=1}^{d^{(\mathtt{mb})}} (\prod_{l=0}^{k_z-1} [1 + l/k_z])}{\prod_{l=0}^{n-1} [1 + l/m]} \label{eq:bosonic_dist}
\end{align}
where, $k_z$ is the bin size which we allow in principle to vary, and $q_z$ = $k_z/m$ is the relative bin size. Ideally one would like to find a lower bound on the average case, i.e. averaged over Haar-random interferometers, TVD between the reference distribution $P^{({\rm mb})}$ and the pre-computable distributions $P_D, P_B$. In the absence of such a bound, we resort to numerics. Fig.~\ref{fig:plot_inset} shows that the TVD between the reference and Haar-averaged distribution is large enough to distinguish between them with a reasonable number of samples for 16 photons and 256 modes. The inset shows a similar trend for different photon numbers.

For the purpose of these simulations, we calculated the mode-binned distribution exactly and did not use Gurvits' approximation as in \cite{SNAC22}. We modified the algorithm to only use $\binom{N+d^{(\mathrm{mb})}-1}{N}$ discrete (non-uniform) points on the grid for the inverse Fourier transform instead of all the $(N+1)^{d^{(\mathrm{mb})}}$ points.  This provides a reduction in the run times by over three orders of magnitude \cite{pow_testnet}. For large $N$, one would need to use Gurvits' approximation because of the exponential complexity of the exact computation of the permanent.

For completeness, we also analyse the minimum number of samples needed to ensure the samples are validated for a particular $\beta$. This means that the TVD between the sample distribution and the exact mode-binned distribution should be less than $\beta$. We find this numerically from the plot of TVD vs. sample sizes. The required number of samples is then plotted against $N$ for various values of $\beta$ and is shown in Fig.~\ref{fig:num_samples_N}. For comparison, we include the theoretical upper bound given by Eq.~\ref{eq:num_tot}. As an example, with $N=16$ and $\beta = 0.02$, Eq.~\ref{eq:num_tot} suggests 14.66 billion samples are needed, whereas in practice, only 3.74 million samples are required. This demonstrates that, in practical scenarios, players need to commit far fewer samples than the theoretical upper bound to pass the validation test.

\subsection{Payoff mechanism}

In Appendix~\ref{payoff}, we show that given some innocuous assumptions, a payoff mechanism can be constructed such that a unique pure strategy Nash equilibrium exists where each player’s dominant strategy is to submit an honest sample from a quantum boson sampler. In other words, we construct a reward and penalty mechanism where it is the player’s unique dominant strategy to behave honestly by committing samples from the boson sampling distribution. This is in contrast to doing nothing, or cheating, i.e. providing samples from some other, presumably easy-to-calculate, distribution. This situation is described by the inequalities, $u_{i}^{honest} > u_{i}^{nothing} > u_{i}^{cheat}$, representing the net utilities of the $i^{th}$ node under the honest, nothing, and cheating strategies respectively. We set a threshold distance $\epsilon$ between the peak bin probability of individual nodes, $\mu_{i}$, and the net PBP $\mu_{net}$, such that if $f(|\mu_i-\mu_{\mathtt{net}}|)<  \epsilon$, the nodes are rewarded in proportion to the number of samples they have committed and otherwise are penalized by losing their initial stake. The function $f$ should be monotonic and we can assume it is linear in the argument. The choice of $\epsilon$ decides the tolerance of the rewarding subroutine and can be changed to incorporate multiple error models.

The assumptions for the payoff mechanism are as follows: ($\it{a}$) a player’s utility is equal to the expected rewards minus the expected penalties and the costs incurred to generate a sample, ($\it{b}$) an individual player's sample contribution is significantly smaller than the combined sample of all players (i.e. $|s_i| \ll |s_{total}|$) so that $\mu_{\mathtt{net}}$ remains unchanged irrespective of $node_i$ being honest or cheating, ($\it{c}$) the verification subroutine is fairly accurate for $|s_i| \gg 1$ so that a honest player will satisfy $f(|\mu_i-\mu_{\mathtt{net}}|) < \epsilon$ with probability $p_i^{\mathtt{honest}} \in \mathbb{R}_{(0.75, 1)}$ and a cheater will satisfy $f(|\mu_i-\mu_{\mathtt{net}}|) < \epsilon$ with probability $p_i^{\mathtt{cheat}} \in \mathbb{R}_{(0, 0.25)}$, ($\it{d}$) the cost to generate sample $\{s_{i}\}$ (denoted $C_i$) scales linearly with $|s_i|$. That is, $C_i = kn$, where $k \in \mathbb{R}$ and $n \equiv |s_i|$. The $k$ parameter here includes costs such as energy consumption to generate one sample but not sunk costs \cite{Bernheim2014} like the initial set-up costs of the classical or quantum devices, ($\it{e}$) the cost to generate a cheating sample is 0. This last assumption provides the most optimistic scenario for a cheater but can be relaxed to accommodate cheating samples with costs. 

Through these assumptions, we establish two results. First, without a penalty, such as that enforced in our protocol by losing the initial staked tokens when failing the validation test, no Nash equilibrium can be established and instead dishonest nodes have no incentive to leave the network. Second, for sufficiently sized rewards, then the dominant strategy is for honest nodes to continue participating, i.e. $u_i^{honest} > 0$, $u_i^{cheat}$, $u_i^{\mathtt{classical}} < 0$, and $u_i^{nothing} = 0$. By the uniqueness of strictly dominant Nash equilibria, this strategy is unique. The criterion on the relative sizes of reward $R$ and penalty $P$ is

\begin{align}
   \frac{R}{3}<P<R, 
\end{align}
where the bound on the reward is
\begin{align}
    2k<R<k^{classical},\label{rewardbound}
\end{align}
where $k^{classical}$ is the per sample cost for a classical node and $k$ is the per sample cost for a node with boson sampler. Note that for a boson sampling distribution of $N$ photons and $M=N^2$ modes, as discussed in Sec.~\ref{SampleRates}, the most efficient known classical boson sampling simulator has a per sample cost exponential in $N$, i.e. $k^{classical} = O(2^N N)$. However, the per sample cost of getting a sample from a boson sampler is linear in $N$, i.e. $k = O(N)$, assuming perfect beam-splitter transmission probabilities ($\eta_t=1$) and photon generation, coupling, and detecting efficiencies ($\eta_f=1$). If $\eta_t,\eta_f<1$ then $k$ increases exponentially with $N$. However, as shown below it can still be many orders of magnitude cheaper than for classical computers for a range of $N$ that is relevant to achieving consensus.

Note that some of these assumptions in this payoff mechanism can be relaxed and still maintain a robust protocol. For example, the split reward system can be replaced with a winner takes all block reward, and capital expenditure $k_{fixed}$ for quantum boson samplers can be accounted for by the replacement $k\rightarrow k_{variable}+k_{fixed}/\tau$ in Eq.~\ref{rewardbound} where $\tau$ is the number of samples the boson sampler is expected to produce before obsolescence.  
\subsection{Quantum vs. classical sampling rates}
\label{SampleRates}

The time needed to successfully mine a block is determined by the inverse of the sampling(repetition) rate of the physical device. For a photonic boson sampler, the repetition rate is \cite{Robens22}
\begin{align}
    R_q=(\eta_f\eta^M_t)^N R_0/N e.\label{Rquantum}
\end{align}
Here $R_0$ is the single photon source rate and $R_0/N$ is the rate at which $N$ indistinguishable photons are produced, $\eta_f$ is a parameter that doesn't scale with the number of modes and accounts for the preparation and detection efficiencies per photon. It can be written as the product $\eta_f=\eta_g\eta_c\eta_d$, where $\eta_g$ is the photon generation efficiency, $\eta_c$ is the coupling efficiency, and $\eta_d$ is the detector efficiency. Finally, $\eta_t$ is the beamsplitter transmission probability. Since we are assuming a circuit of depth equal to the number of modes (which is sufficient to produce an arbitrary linear optical transformation), the overall transmission probability per photon through the circuit is $\eta_t^M$. Finally, the factor of $e$ is an approximation of the probability of obtaining a collision-free event \cite{AC11}. The experiment of Ref.~\cite{PhysRevLett.123.250503} produced a single photon repetition rate of $R_0=76$MHz and the experiment of Ref.~\cite{deng2023gaussian}, reported a transmission probability per photon through a $144\times 144$ optical circuit of $97\%$ implying a per beamsplitter transmission probability of $\eta_t=0.97^{1/144}$ as well as an average wavepacket overlap of $99.5\%$, demonstrating high indistinguishability. A photon generation efficiency of $\eta_g=0.84$ was reported for quantum dot sources in Ref.~\cite{Arcari2014} and efficiencies of $\eta_c=0.9843$ have been demonstrated for coupling single photons from a quantum dot into a photonic crystal waveguide \cite{Arcari2014}. Finally, single photon detector efficiencies of up to $\eta_d=0.98$ have been reported at telecom wavelengths \cite{You+2020+2673+2692}.
All these numbers can reasonably be expected to improve as technology advances \cite{bib:ChinaBSRev, alexander2024manufacturableplatformphotonicquantum}.


The state-of-the-art general-purpose method to perform classical exact boson sampling uses a hierarchical sampling method due to Clifford \& Clifford \cite{CC2018}. The complexity is essentially that of computing two exact matrix permanents providing for a repetition rate \footnote{We ignore the relatively small $O(MN^2)$ additive complexity to the classical scaling.} 
\begin{align}
    R_c = \frac{1}{\tilde{a}\cdot 2 \cdot N \cdot 2^N}.
\label{Rclassical}
\end{align}
Here $\tilde{a}$ refers to the scaling factor (in units of seconds $s$) in the time to perform the classical computation of the matrix permanent of one complex matrix where Glynn's formula is used to exactly compute the permanent of a complex matrix in a number of steps $O(N 2^N)$ using a Gray code ordering of bit strings.
Recently an accelerated method for classical boson sampling has been found with an average case repetition rate scaling like $R_c=O(1.69^{-N}/N)$ \cite{clifford2020faster}, however, this assumes a linear scaling of the modes with the number of photons, whereas we assume a quadratic scaling.

\begin{figure}[t!]
\includegraphics[width=\columnwidth]{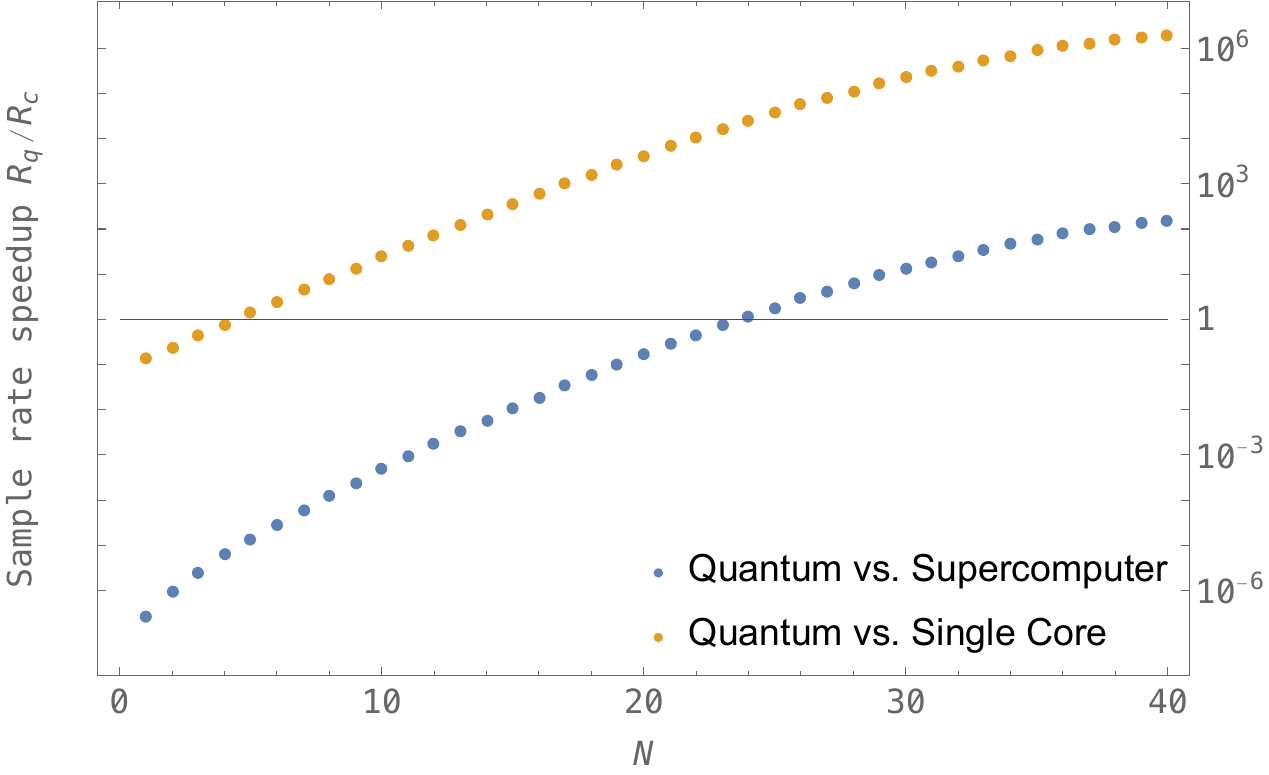}
    \caption[Caption for LOF]{Sampling rate speed up $R_q/R_c$ (log scale) for quantum boson-samplers relative to classical computers. Points above the red line indicate a quantum speedup. Orange dotted line: performance relative to single-core Intel Xeon processor running at 3.5GHz, and 128GB RAM with $\tilde{a}=10^{-9.2}$s \cite{LUNDOW2022110990}. Blue dotted line: performance relative to the Tianhe-2 supercomputer \cite{10.1093/nsr/nwy079}  with $\tilde{a}=N\times 1.99\times 10^{-15}$s\footnotemark. The photonic boson-sampler is assumed to have the following specifications: single photon source rate $R_0=100$MHz, a single photon joint preparation and detection probability of $\eta_f=0.90$, and beamsplitter transmission probability of  $\eta_t=0.9999$.}
    \label{fig:analysis2}
\end{figure}

As shown in Fig.~\ref{fig:analysis2} the performance ratio, defined as the ratio of sampling rates for quantum to classical machines $R_q/R_c$, is substantial even for a modest number of photons.


\footnotetext{The extra factor of $N$ comes from the fact that the parallelization here cannot directly implement Gray code ordering in the Balasubramanian–Bax–Franklin–Glynn algorithm (a variant of Ryser's algorithm described in App.~\ref{appendix:modebinest}). The quoted number is for the calculation done in 2018 using all available $16000$ nodes on the supercomputer, each containing three CPUs and two co-processors.}

\begin{figure}[t!]
\includegraphics[width=\columnwidth]{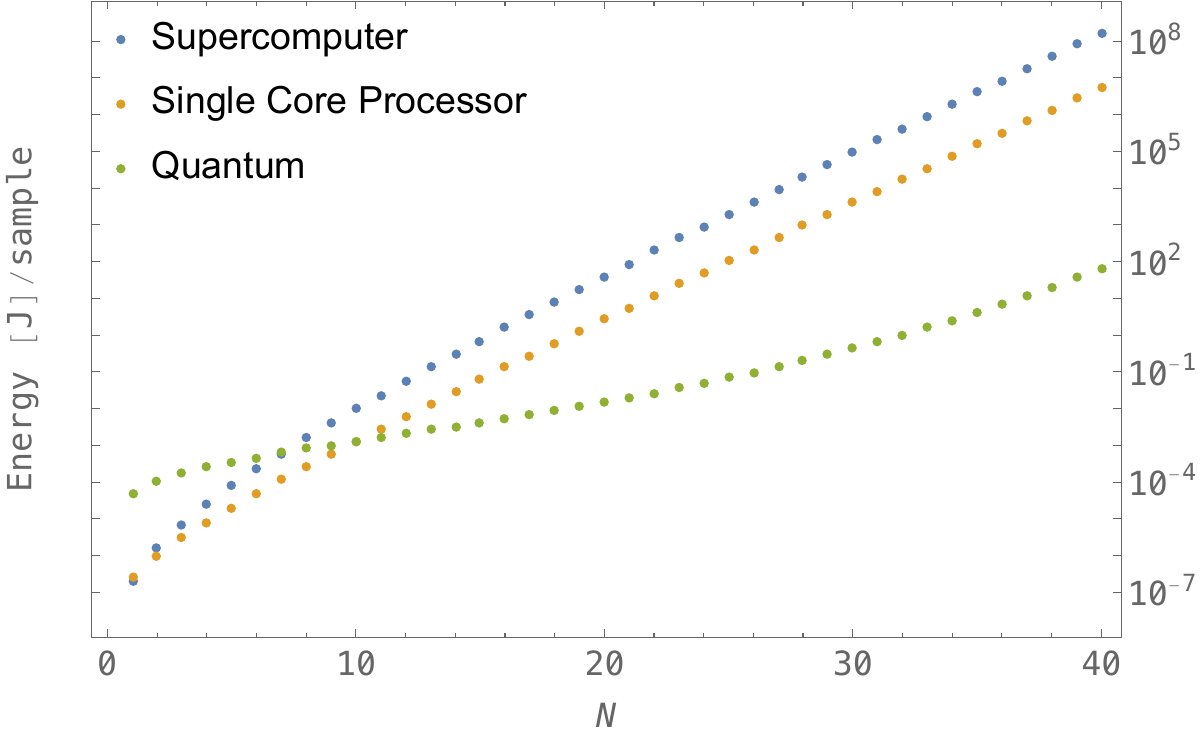}
    \caption{Comparison of the energy cost per sample (log scale) for boson-sampling using: a quantum boson-sampler, a supercomputer, and a single core processor all with same specs as in Fig.~\ref{fig:analysis2}.}
    \label{fig:analysis3}
\end{figure}

\subsection{Quantum vs. classical energy cost}

The energy cost to run boson samplers is dominated by the cost to cool the detectors and the single-photon sources. Superconducting single-photon detectors of NbN type with reported efficiencies of $\eta_d=0.95$ can operate at $2.1$K \cite{You+2020+2673+2692}, which is just below the superfluid transition temperature for Helium, and semiconductor quantum dot sources can run at high efficiency and indistinguishability below $10$K \cite{iles2017phonon}.  Two-stage Gifford-McMahon cryocoolers can run continuously at a temperature of $2$K with a power consumption of $\sim 1.5$kW \cite{You+2020+2673+2692}.  
To compare the energy cost of boson-samplers to classical samplers, note that the power consumption of the Tianhe-2 supercomputer is $24$MW \cite{10.1093/nsr/nwy079}, and the power consumption of a single core processor at $3.5$GHz is $\sim 100$W. Ultimately, the important metric is the energy cost per sample since it is the accumulation of valid samples that enables a consensus to be reached. As seen from Fig.\ref{fig:analysis3}, quantum boson-samplers are significantly more energy efficient than classical computers. For example, at $N=25$ photons the quantum boson-sampler costs $6.77\times 10^{-2}$J per sample which is $1563$ times more energy efficient than a single core processor and $29569$ times more efficient than a supercomputer.

While classical devices, such as ASICs, could be developed in the future that would speed up calculations of matrix permanents by some constant factor, any such device is fundamentally going to be limited by the exponential in $N$ slowdown in the sampling rate ($R_c$ in Eq.~\ref{Rclassical}). Even as classical computers do speed up, one can increase the number of photons to maintain the same level of quantum advantage. Importantly, this would not require frequent upgrades on the boson sampler since the same device can accommodate a few more input photons as the number of modes was already assumed to be $O(N^2)$. Furthermore, as the quality of the physical components used for boson sampling improves, the quantum repetition rates ($R_q$ in Eq.~\ref{Rquantum}) will increase, ultimately limited by the single photon source rate. 

On the other hand, it is unlikely that much faster ``quantum ASIC'' devices will be developed for boson sampling. Fock state Boson sampling can be simulated fault tolerantly by universal quantum computers with polynomial overhead. One way to do this is to represent the state space as a truncated Fock space encoded in $M$ qudits of local dimension $N+1$ (or in $M\times\lceil \log{(N+1)}\rceil$ qubits). The input state is a tensor product state of $\ket{0}$ and $\ket{1}$ states, the gates of the linear interferometer are two qudit gates which can be simulated in $O(N^4)$ elementary single and two qudit gates, and the measurement consists of local projectors such that the total simulation scales like $O(N^4 M^2)$. Another way to translate boson sampling to quantum circuits performs a mapping to the symmetric space of qudits as described in \cite{MoylettTurner2018}. Given the algorithmic penalty as well as the gate overheads for error correction, the quantum computer simulation would be slower than a photonic-based native boson sampler, except in the limit of very large $N$ where the fault tolerance of the former enables a speedup. However, at that point, the entire protocol would be too slow to be of use for consensus anyway. 

The improvements in the quantum repetition rates will hinge on advances in materials and processes that most likely would impose a negligible increase in energy cost. In this sense, PoW by boson sampling offers a route to reach a consensus without inducing users to purchase ever more power-hungry mining rigs. 

\section{Discussion}

We have proposed a PoW consensus protocol that natively makes use of the quantum speedup afforded by boson samplers. The method requires that miners perform full boson sampling, where samples are post-processed as coarse-grained boson sampling using a binning strategy only known after samples have been committed to the network. This allows efficient validation but resists pre-computation either classically or quantum mechanically.

Whereas classical PoW schemes such as Bitcoin's are notoriously energy inefficient, our boson-sampling-based PoW scheme offers a more energy-efficient alternative when implemented on quantum hardware. The quantum advantage has a compounding effect: as more quantum miners enter the network the difficulty of the problem will be increased to maintain consistent block mining time, further incentivizing the participation of quantum miners.


The quantum hardware required for the implementation of our protocol has already been experimentally demonstrated at a sufficient scale and is becoming commercially available.
While we have focused our analysis primarily on conventional Fock state boson sampling, the method extends to Gaussian boson sampling, accommodating faster quantum sampling rates owing to the relative ease with which the required squeezed vacuum input states can be prepared. We have used conservative estimates of the number of samples needed to achieve consensus and it is plausible that these upper bounds can be tightened. We leave a detailed study of the number of samples required, error tolerances, and performance of Gaussian boson samplers to future work.

Like the inverse hashing problem in classical PoW, the boson-sampling problem has no intrinsic use. It would be interesting to consider if samples contributed to the network over many rounds could be used for some practical purpose, enabling `useful proof-of-work', something that has also been suggested in the context of conventional blockchains \cite{e21080723}.



\section{Methods} \label{sec:Methods}

In this section, we will provide detailed descriptions of the various steps involved in the protocol. For an illustration of these steps, please refer to Fig.~\ref{fig:consensus}.

\begin{enumerate}
\item A transaction, or bundle of transactions, is created on the network. All nodes are aware of the following set of input parameters:
\begin{align}
{\tt Pm}=\{N, M, U, d^{(\mathtt{mb})},d^{(\mathtt{sb})},T_{\mathtt{mine}},\epsilon,\beta ,R, P\},
\end{align}
which is assumed to be constant over many blocks but can be varied to adjust the difficulty of the problem.
\item A new block $b_j$ representing this transaction is created. It has a header ${\tt header}(b_j)$ that
contains summary information of the block including the parameter set ${\tt Pm}$, a hash derived from transactions in the block, a hash of the previous block header together with its validation record ${\tt Rec}(b_{j-1})$ (discussed in step 7), and a timestamp.  
\item The new block is sent to every node in the network. All nodes stake tokens to participate. Note this is different from a proof-of-stake protocol since here all miners stake the same amount of tokens and the probability of successfully mining a block is independent of the staked amount. 
\item Miners implement boson-sampling \cite{bib:aa} using devices like those illustrated in Figure \ref{fig:boson_sampling_permutation}, using $N$ photons input into $M$ modes ordered $\{1,2,\ldots ,M\}$. A hash of the header is mapped to a permutation on the modes using a predetermined function $a$
\begin{align}
    a: H({\tt header}(b_j))\to \Pi\in S_M.
\end{align}
This permutation, which depends on the current block, is used to determine the locations of the $N$ input photons in the input state of the boson sampler. 
Each node $i$ collects a set of samples denoted $s_i$, of size $|s_i|$, and commits each sample in the set by hashing that sample along with a timestamp and some private random bit string. The committed samples are broadcast to the network. The set of committed samples by node $i$ is denoted $\tilde{s}_i$. The purpose of broadcasting hashed versions of the samples is to prevent dishonest miners from simply copying honest miners' samples. 
\item After some predetermined mining time
\begin{align}
    T_{\rm mine} = \max\{N_{tot}^{(\mathtt{mb})},N_{tot}^{(\mathtt{sb})}\}/R_q,
\end{align}
where $R_q$ is an estimated quantum sample rate defined in \eqref{Rquantum}, the mining is declared over and no new samples are accepted. Note that $R_q$ can be periodically updated to account for changes in network sample production akin to difficulty parameter adjustments in other PoW protocols \cite{conf:nakamoto}. All miners then reveal their sample sets $\{s_i\}$ as well as the random bit strings associated with each sample so that the sets can be verified against the committed sets $\{\tilde{s}_i\}$. If for some node $i$, the sets don't agree, that node is removed from further consideration of the mining round and they lose their stake.
Let the set of remaining samples be $W=\bigcup_i s_i$.

\item 
This stage consists of three steps: a validation step using mode binning to catch dishonest miners, a state binning step to determine the mining success criterion, and a reward/penalty payoff step.
\begin{enumerate}
    \item {\it Validation}. A mode-binned distribution $P^{(\mathtt{mb})}$ is used to validate each miner's sample set. Mode binning refers to grouping output modes into $d^{(\mathtt{mb})}$ bins so that for a given sample the number of photon counts in a bin is simply the total number of ones at all the bit locations contained in the bin. We assume the bins are of equal size
\begin{align}
|\mathtt{bin}^{(\mathtt{mb})}_j|=M/d^{(\mathtt{mb})}\ \forall j.
\end{align}
     A random beacon in the form of a string ${\tt beacon}^{(\mathtt{mb})}$ is announced to the network. Decentralized randomness beacons can be integrated into PoW consensus protocols in such a way that they are reliable, unpredictable, and verifiable. It would be advisable here to construct the beacons using post-quantum secure verifiable random functions \cite{li2021postquantum, Buser}. Using a predetermined function $g$, 
    \begin{align}
        g:{\tt beacon}^{(\mathtt{mb})}\rightarrow \pi^{(\mathtt{mb})}\in S_{M},
      \end{align}
    The beacon is mapped to a permutation on the modes
    such that the modes contained in \mbox{$\mathtt{bin}^{(\mathtt{mb})}_j$} are
    \begin{align}
        \{\pi^{(\mathtt{mb})}(k)\}_{k=(j-1)M/d^{(\mathtt{mb})}+1}^{jM/d^{(\mathtt{mb})}}.
    \end{align}
      The mode-binned distribution for miner $i$ is
      \begin{align}
P^{(\mathtt{mb})}[i] (\mathbf{n})=\frac{\omega[i](\mathbf{n})}{|s_i|}  ,
      \end{align}
     where $\omega[i](\mathbf{n})$ is the number of times the binned multiphoton configuration $\mathbf{n}$, as defined in Eq.~\ref{ouputbinvec}, has occurred in the sample set $s_i$ committed by the $ i^{th }$ miner. Note that for $N$ photons in $d^{(mb)}$ bins, a total of ${d^{(\mathtt{mb})} + N - 1}\choose{N}$ configurations are possible. The true mode binned distribution, $P^{(\mathtt{mb})}$, that depends on \mbox{$(\Pi,\pi^{(\mathtt{mb})}, U)$}, can be estimated as $\widehat{P^{(\mathtt{mb})}}$ using a polynomial time classical algorithm. If the total variation distance between the distributions \mbox{$\mathcal{D}^{(tv)}(\widehat{P^{(\mathtt{mb})}},P^{(\mathtt{mb})}[i])\geq 2 \beta$} for some predetermined \mbox{$0<\beta<1$} then the sample set $s_i$ is invalidated and miner $i$ loses their stake. Otherwise, the sample set is validated and labeled $s_i^{(v)}$. Let the set of validated samples be
     \begin{align}
         W^{(v)}=\bigcup_i s^{(v)}_i.
        \end{align}
    \item
    {\it Determining success criterion}.  At this step a state binned distribution $P^{(\mathtt{sb})}$ is computed to determine which miners are successful. First, it is necessary to sort the samples in $W^{(v)}$ into bins, a procedure referred to as state binning.  The state space $Y$ consists of $(N+1)$ary valued strings of length $M$ and weight $N$:
    \begin{align}
        Y=\{Y_k\}=\{(y_1^{(k)},\dots,y_M^{(k)});\nonumber\\
    y_j^{(k)}\in\mathbb{Z}_{N+1},\sum_{j=1}^M y_j^{(k)}=N\},
    \end{align}
    where the notation $y_i^{(k)}$ means for the $k^{th}$ element of the sample space $y_i$ photons were measured in the $i^{th}$ mode. The states in $Y$ are ordered lexographically\footnote{For example, for \mbox{$M=3, N=2$} the ordering would be
\mbox{$\{(002),(011),(020),(101),(110),(200)\}$}.}. 
A second beacon ${\tt beacon}^{(\mathtt{sb})}$ is announced to the network and using a predetermined function $f$,
\begin{align}
    f:{\tt beacon}^{(\mathtt{sb})}\rightarrow \pi^{(\mathtt{sb})} \in S_{|Y|}.
\end{align}
The beacon is mapped to a permutation on the state space.
The states are sorted into $d^{(\mathtt{sb})}$ equal sized bins such that the states contained in $\mathtt{bin}^{(\mathtt{sb})}_j$ are
\begin{align}
\{Y_{\pi(k)}\}_{k=(j-1)|Y|/d^{(\mathtt{sb})}+1}^{j|Y|/d^{(\mathtt{sb})}}.
\end{align}
All the publicly known samples in $W^{(v)}$ are then sorted into the bins and the collective state binned distribution is
\begin{align}
    P^{(\mathtt{sb})}=\frac{1}{|W^{(v)}|}(h_1,h_2,\ldots, h_d),
\end{align}
where $h_j$ is the number of samples in the $\mathtt{bin}^{(\mathtt{sb})}_j$. The PBP across the validated miners in the network is
\begin{align}
    \mu_{\rm net}=\frac{\max_j\{h_j\}}{|W^{(v)}|}.
\end{align}
Similarly, the PBP for validated miner $i$ is
\begin{align}
\mu_i=\frac{\max_j\{|s^{(v)}_i\cap bin_j|\}}{|s^{(v)}_i|}.
\end{align}
    \item 
{\it Payoff}. Miners whose samples were validated have their stake returned and are awarded a payoff if $|\mu_i-\mu_{\rm net}|\leq \epsilon$ for some predetermined precision $\epsilon$. The amount of the payoff is dependent on the number of samples committed. 
\end{enumerate}
\item The new block $b_j$ is added to the blockchain with an appended record
\begin{align}
{\tt Rec}(b_j)=\{\Pi,\pi^{(\mathtt{mb})},\pi^{(\mathtt{sb})},\widehat{P^{(\mathtt{mb})}},\mu_{\rm net}\}.
\end{align}
This record contains the information necessary to validate the block. 
\end{enumerate}

\subsection{Variation of the protocol using Gaussian Boson-Sampling} \label{Sec:GBS}

While the original boson-sampling protocol described was based on photon-number states, variants based on alternate types of input states have been described \cite{bib:olson_sq_bs, bib:kaushik_bs}. Most notably, Gaussian boson-sampling (GBS) \cite{bib:gaussian_bs, RevModPhys.84.621}, where inputs are Gaussian states (specifically squeezed vacuum states), has gained a lot of traction amongst experimental realizations owing to the relative ease and efficiency of preparing such states. A variation of the whole protocol can be done using coarse grained Gaussian boson sampling. A classical verification scheme using the characteristic function, similar to the method described in \cite{SNAC22}, can be constructed. In this case, the output will be post-selected on a fixed total photon number subspace. A slight modification of the scheme enables one to exactly calculate the characteristic function for the case of GBS (see Sec.~ \ref{appendix:CharGBS} for details).

Here we discuss how Gaussian boson-sampling can be used in place of Fock-state boson-sampling for the scheme. Many of the existing protocols for photon generation were already making use of Gaussian states and post-selection, so the complexity of sampling from the output state when the input state is a Gaussian state was studied in detail \cite{bib:gaussian_bs}. Gaussian states can be characterized by their mean and variance. The simplest Gaussian states are coherent states. It is interesting to note that there is no quantum advantage in using coherent states as input states for boson sampling. In this variant of boson sampling, input states are taken to be squeezed vacuum states. The squeezing operator is given by
\begin{align}
    \hat{S}(z) = \exp\left[\frac{1}{2} ( z^* \hat{a}^2 - z \hat{a}^{\dagger 2} )\right],\,\, z=re^{i\theta}.
\end{align}

Let us assume a Gaussian Boson-Sampling setup with squeezed vacuum states in $N$ of $M$ modes and vacuum in the remaining $M-N$ modes. The initial state is
\begin{align}
    \ket{\psi_{\rm{in}}} = \prod_{j=1}^N \hat{S}_j(r_j) \ket{0}, \label{eq:GBS_in}
\end{align}
where $r_j$ is the squeezing parameter for the $j$th mode, which is assumed to be real for simplicity. 
The symplectic transformation corresponding to the squeezing operations is 
\begin{align}
    S = \begin{pmatrix}
\oplus_{j=1}^M \cosh{r_j} & \oplus_{j=1}^M \sinh{r_j}\\
\oplus_{j=1}^M \sinh{r_j} & \oplus_{j=1}^M \cosh{r_j}
\end{pmatrix}. 
\end{align}
Then the covariance matrix for the output state after the input state passes through the interferometer described by $U$ is 
\begin{align}
\sigma = \frac{1}{2} \begin{pmatrix}
U & 0\\
0 & U^*
\end{pmatrix} SS^{\dagger}\begin{pmatrix}
U^{\dagger} & 0\\
0 & U^T.
\end{pmatrix}.
\end{align}
Now let the particular measurement record of photon number counts be $Y_k=(y^{(k)}_1,\ldots, y^{(k)}_M)$. Then the probability of finding that record is given by
\begin{align}
    \rm{Pr}(Y_k) &= |\sigma_{Q}|^{-1/2} |\rm{Haf}(B_{Y_k})|^2, \nonumber \\
    \sigma_Q &= \sigma + \frac{1}{2} \mathbb{I}_{2M}. \label{eq:sigmaq}
\end{align}
Here the matrix $B_{Y_k}$ is a constructed from the matrix 
\begin{align}
    B = U(\oplus_{j=1}^M \tanh{r_j}) U^T,
\end{align}
and 
is determined as follows. If $y_i=0$ then rows and columns $i$ of matrix $B$ are deleted, otherwise the rows and columns are repeated $y_i$ times.
Haf($\cdot$) denotes the matrix Hafnian. Similar to the permanent, the Hafnian of a general matrix is also \textbf{\#P}-hard to calculate. It has been shown that sampling from the output state is also hard in the case of Gaussian boson sampling.

We can think of analogous mode and state-binned sampling for the Gaussian variant. In both cases, the output state is post-selected onto a fixed total photon number subspace. The probability of this can be absorbed to the repetition rate. For the mode-binned Gaussian boson sampling we will want to develop a validation scheme similar to the one described in \cite{SNAC22}. Even though other methods exist for validating samples from Gaussian boson sampling \cite{bib:DrummondBS}, we would like to have a protocol similar to that was used for Fock-state boson sampling. The detailed study of the parameters involved including the required number of samples required is beyond the scope of this paper. 

The protocol is similar to Sec.~\ref{sec.mode-bin} in the main text. We start with the input state defined in Eq.~\ref{eq:GBS_in}. The squeezing parameter is taken so that the total average number of photons is close to $2N$. Then the probability, $P^{({\mathtt{mb}})}({\bf n})$, of measuring the binned output configurations can be expressed as 
\begin{align}
    P^{({\mathtt{mb}})}({\bf n})=\frac{1}{(N+1)^{d^{(\mathtt{mb})}}}\sum_{{\bf \tilde{c}}\in \mathbb{Z}_{N+1}^{d^{(\mathtt{mb})}}}\tilde{\chi}\left(\frac{2\pi {\bf \tilde{c}}\cdot {\bf n}}{N+1}\right)e^{-i \frac{2\pi {\bf \tilde{c}}\cdot {\bf n}}{N+1} }.
\label{eq:probmb}
\end{align}

The calculation of the characteristic function is slightly different since now the input state does not have a fixed number of photons. It is as follows
\begin{align}
    \chi(\mathbf{c}) &= \sum\limits_{n_k|k=1,2,\cdots m} P({\bf n}) e^{i \bf c.\bf n}. \label{eq:charfunc}
\end{align}
It was shown in Ref.~\cite{Kocharovsky22} (see Eq.~25 within reference) that the characteristic function for GBS is

\begin{align}
    \chi(\mathbf{c}) &= \frac{1}{\sqrt{\det\left(\mathbb{I} - Z(\mathbb{I} - \sigma_Q^{-1})\right)}},\\
    Z &=\bigoplus\limits_{k=1}^M \begin{bmatrix}
    e^{i\frac{2\pi c_k}{N+1}} & 0 \\
    0 & e^{i\frac{2\pi c_k}{N+1}}
\end{bmatrix}.
\end{align}
Here $\sigma_Q$ is related to the covariance matrix of the output state and is defined in Eq.~\ref{eq:sigmaq}, and  $\tilde{\chi}\left({\bf \tilde{c}}\right)$ can be obtained from $\chi(\mathbf{c})$ by replacing all $c_k$'s in $i^{\text{th}}$ bin to be $\tilde{c}_i$ (see Appendix \ref{appendix:CharGBS} for more details). This function can now be used in Eq.~\ref{eq:probmb} and evaluated at a polynomial number of points to obtain the exact binned distribution (see also Ref.~\cite{Drummund2022} for an alternative approach using classical sampling of the positive P distribution to obtain an approximation of the mode binned distribution). The rest of the protocol is similar to that of Fock state boson sampling. 

\acknowledgements{We gratefully acknowledge discussions with Louis Tessler, Simon Devitt, Peter Turner, Jelmer Renema, and Sanaa Sharma. GKB and GM received support from a BTQ-funded grant with Macquarie University. GKB and PPR receive support from the Australian Research Council through the Centre of Excellence for Engineered Quantum Systems (project CE170100009). DS is supported by the Australian Research Council (ARC) through the Centre of Excellence for Quantum Computation and Communication Technology (project CE170100012).} 


\bibliography{main} 
\appendix

\section{Supplementary information}

\subsection{Blockchains}

\begin{figure}[t!]
    \includegraphics[width=\columnwidth]{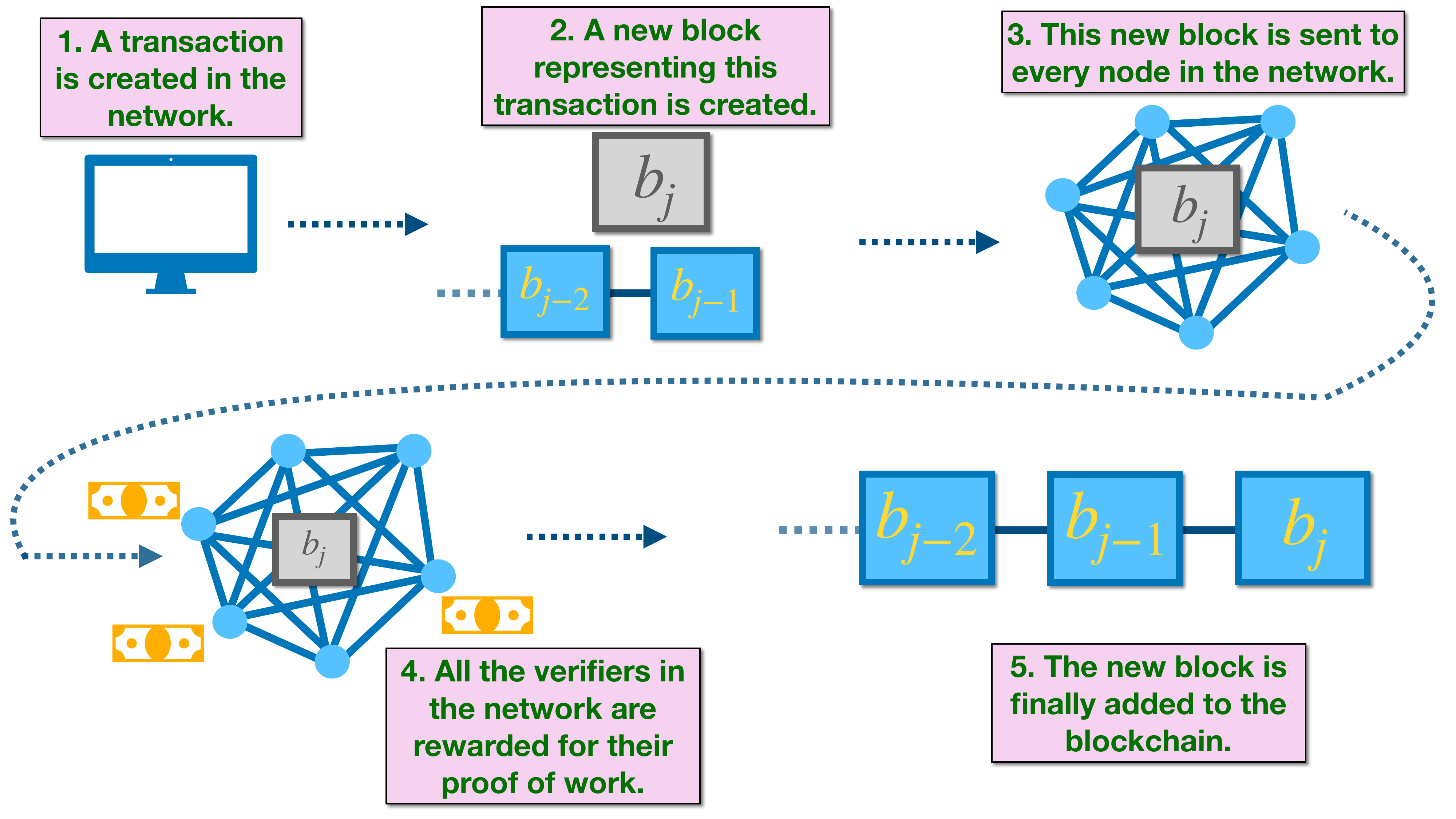}
    \caption{Blockchain architecture and the addition of new blocks.}
    \label{fig:process}
\end{figure}

A blockchain is a decentralized and distributed ledger that stores transactions in a secure and transparent manner. The ledger consists of a chain of fixed-length blocks, each of which is verified by every node in the network. The network is decentralized, meaning no central authority exerts control, relying on a network of nodes to maintain its integrity. Each block is then added to the blockchain once a decentralized consensus is reached. The whole process is illustrated in Fig.~\ref{fig:process} and can be described as follows:
\begin{enumerate}
    \item Transaction Verification: Transactions are sent to the network. Before a transaction can be included in a block, it must be validated by nodes on the network. Each node checks that the transaction is legitimate and that the sender has sufficient funds to complete the transaction.
    \item Block Creation: Once a group of transactions is verified, they are bundled together into a block. The block contains a header, which includes the previous block's hash, a timestamp, and a \emph{nonce} (a random number).
    \item Proof-of-Work: To mine the block, miners compete to solve a complex mathematical puzzle, known as Proof-of-work (PoW). The first miner to solve the puzzle broadcasts their solution to the network, and the other nodes verify the solution. If the solution is correct, the miner is rewarded with newly minted cryptocurrency, and the block is added to the blockchain.
    \item Consensus Mechanism: To maintain the integrity of the blockchain, the network must reach a consensus on the state of the ledger. In a decentralized blockchain network, this is achieved through a consensus mechanism, such as PoW or Proof-of-Stake (PoS). PoW requires miners to compete to solve a mathematical puzzle, while PoS relies on validators who hold a stake in the network to verify transactions.
    \item Block Confirmation: Once a block is added to the blockchain, it cannot be altered or deleted. Other nodes on the network can confirm the block by verifying the hash of the previous block, ensuring that the chain is continuous and secure.
\end{enumerate}

\subsubsection{One-way functions}

Blockchain technology relies heavily on one-way functions as a critical component of its security infrastructure. One-way functions are mathematical functions that are easy to compute in one direction but difficult to reverse. That is, given a function $f(x) = y$, $y$ is easy to compute for all inputs $x$, however, computing $x$ for a given $y$ is hard. Computationally speaking, the notions of `easy' and `hard' refer to polynomial-time and super-polynomial-time algorithms respectively in the input size. Therefore, in general, the inversion of one-way functions resides within the computational complexity class \textbf{NP} (efficiently classically verifiable) since the verification of any pre-image is possible in polynomial time, unlike its explicit computation. 

The public-key cryptography used in blockchains today relies on pairs of related keys (public and private) generated by one-way functions. While it is easy to compute a public key from a private key, the reverse operation is computationally intractable. This makes private keys extremely difficult to guess or brute-force, thus ensuring the security of blockchain networks. 


\subsubsection{Hash functions}

A general hash function is a one-way function that satisfies three main properties: $\it(a)$ its input can be of any size, $\it(b)$ its output is always of a fixed size, and that $\it(c)$ it should be easy to compute. 
A cryptographic hash function $H(x)$ has several additional requirements including collision-freeness, hiding, and puzzle friendliness \cite{10.5555/2994437}. 

In some existing classical blockchain implementations, notably Bitcoin \cite{conf:nakamoto}, partial inverse hashing is employed for the purposes of PoW. Here the miners compete to find bitstrings that hash to an output string with some number of leading zeros. The number of required leading zeroes translates to the difficulty of solving this problem. Since hash functions are highly unstructured, the best classical approach to finding such solutions is using brute force to hash random input strings until by chance a satisfying output is found. Once found, it is trivial for other nodes to verify the solution by simply hashing it.

Proof-of-work can be generalised to quantum proof-of-work by considering problems that are both in \textbf{NP} and \textbf{BQP} (efficiently solvable on a quantum computer) but not in \textbf{P} (efficiently classically solvable). Examples of this include integer factorisation and the discrete logarithm problem, which can be efficiently solved on a quantum computer using Shor’s algorithm, and can be efficiently classically verified, but cannot be efficiently classically solved. This would incentivize potentially more energy-efficient quantum mining, however, it would require fault-tolerant quantum computers, devices that are some years away and would likely be expensive to access, and also it is not clear that such functions would satisfy the desirable progress-free condition described above. 

The alternative we describe here is to use non-universal quantum devices that solve the boson sampling problem for PoW. Unlike the one-way functions described above, classical or quantum verification of boson sampling is not known to be possible and hence it would appear hopeless as a means to reach consensus. However, by coarse-graining the distribution, verification indeed is possible.  
State-binned boson-sampling (see Sec.~\ref{sec:sbbs}) was motivated as an attempt to construct a hash function from the boson-sampling problem \cite{Nikolopoulos2019}. This type of function determined by sampling differs from conventional hash functions as it is not in \textbf{NP}, since a classical verifier cannot efficiently verify the output to the hash given the input state. However, binned distributions converge with sufficient samples, enabling verification by quantum verifiers by comparing for consistency in binned distributions.
Moreover, using a different kind of course graining by binning output modes, an efficient classical verification can be achieved. We combine both CGBS strategies in the full protocol to check that miners are producing valid samples and to provide rewards to successful miners in a manner that is responsive to the combined sampling power of the network.

\subsection{Time complexity of estimation of matrix permanent using Gurvits algorithm} \label{appendix:modebinest}

An additive approximation of the permanent of a matrix with complex entries was proposed by Gurvits \cite{gurvits_deterministic_2002, aaronson_gurvits_2012}. It follows by writing the permanent of a matrix $A\in \mathbb{C}^{n\times n}$ as the expectation value of a so-called Glynn estimator $Gly_{x_i}(A)$ over $n$ bit random variables $x_i=(x_{i_1}x_{i_2}\ldots x_{i_n})$ with $x_{i_j}\in \{-1,1\} \forall j \in [n]$:
\begin{align}
    \mathrm{Per}(A)=E_{x_i\in\{-1,1\}^n}[\mathrm{Gly}_{x_i}(A)],
\end{align}
where,
\begin{align}
    \mathrm{Gly}_{x_i}(A)=x_{i_1}x_{i_1}\ldots x_{i_n} \prod_{j=1}^n (A_{j,1}x_{i_1}+\cdots +A_{j,n}x_{i_{n}}).
\end{align}
Importantly, the Glynn estimator can be upper-bounded in its absolute values as follows: 
\begin{align}
    \vert \mathrm{Gly}_{x_i}(A) \vert \leq \Vert A \Vert^{n},
\end{align}
where $\Vert A \Vert$ is the spectral norm of the $A$ matrix, i.e. its largest singular value. 

The algorithm to approximate the permanent of any such matrix $A$ is then as follows:
\begin{enumerate}
    \item Pick $m$ number of $n$-bit strings: $x_1,x_2,\dots, x_m\in \{ -1,1\}^n $.
    \item For each $i \in [m]$, compute $\mathrm{Gly}_{x_i}(A)$. The average of these $\widehat{\mathrm{Per}(A)}=\frac{1}{m}\sum_{i=1}^m \mathrm{Gly}_{x_i}(A)$ is then the estimate of $\mathrm{Per}(A)$.
\end{enumerate}

Using Hoeffding's inequality, one can use the upper bound on $\vert Gly_{x_i}(A) \vert$ to write the following concentration bound for the mean of random Glynn estimators for some $\lambda > 0$:
\begin{align}
   {\rm Pr}\big( \vert \widehat{\mathrm{Per}(A)} - \mathrm{Per}(A) \vert \geq \lambda) \leq 2 e^{-\frac{m \lambda^{2}}{2 \Vert A \Vert^{2n}}}.
\end{align}

Setting $\lambda = \delta \Vert A \Vert^n$, one can then write,
\begin{align}
    {\rm Pr}\big( \vert \widehat{\mathrm{Per}(A)} - \mathrm{Per}(A) \vert < \delta \Vert A \Vert^n) \geq 1 - 2 e^{-\frac{m \delta^{2}}{2}}. 
\end{align}

Therefore, $m = O(1/\delta^2)$ samples suffice to approximate $\mathrm{Per}(A)$ within $\pm \delta \Vert A \Vert^n$ additive error with high probability. Since for each random string $x_i \in \{-1,1\}^n$, the Glynn estimator can be computed in $O(n^2)$ time, the complete algorithm for permanent approximation can be run in $O(n^2/\delta^2)$ time. 

Moreover, if $A$ is a unitary matrix or a sub-matrix thereof,  $\Vert A \Vert \leq 1$. Hence each point in the characteristic function $\chi({\bf s})=\mathrm{Per}(V_N({\bf s}))$ can with probability at least $p$ be computed to within additive error $\delta$ in time,
\begin{align}
    O\left(\frac{N^2\ln(2/(1-p))}{\delta^2}\right).
\end{align}

\subsection{Characteristic function for Gaussian Boson Sampling} \label{appendix:CharGBS}

In order to develop a validation protocol for Gaussian Boson Sampling similar to that for the Fock state, we need to calculate the corresponding characteristic function. As before, the characteristic function is defined as,

\begin{align}
    \chi(\mathbf{c}) &= \sum\limits_{n_k|k=1,2,\cdots m} P({\bf n}) e^{i \bf c.\bf n}. 
\end{align}
We have a closed form for this characteristic function that was calculated in Ref.~\cite{Kocharovsky22} (see Eq.~25 within reference) 
\begin{align}
    \chi(\mathbf{c}) &= \frac{1}{\sqrt{\det\left(\mathbb{I} - Z(\mathbb{I} - \sigma_Q^{-1})\right)}} \\
    Z &=\bigoplus\limits_{k=1}^M \begin{bmatrix}
    e^{ic_k} & 0 \\
    0 & e^{ic_k}.
\end{bmatrix}
\end{align}
Here $\sigma_Q$ is related to the covariance matrix of the output state and is defined in Eq.~\ref{eq:sigmaq}. 

The inverse Fourier transform to obtain $P({\bf n})$ is as follows,
\begin{align}
    P({\bf n}) &= \int_{-\pi}^{\pi}\int_{-\pi}^{\pi} \cdots \int_{-\pi}^{\pi} \chi(\mathbf{c}) e^{-i \mathbf{c}.\mathbf{n}} {\prod_{j=1}^M} \frac{dc_j}{2\pi}.
\end{align}

We would like to calculate the binned distribution $P^{({\rm mb})}(\tilde{\bf n})$ instead of $P(\bf n)$.

\begin{align}
    P^{({\rm mb})}(\tilde{\bf n}) &= \sum_{\bf n| \sum\limits_{i \in \mathrm{bin}_k} n_i = \tilde{n}_k}   P(\bf n).
\end{align}
We can get this by manually introducing some delta functions in the inverse Fourier transform. We define $\tilde{\bf c}$, such that $\tilde{\bf c}_k = c_i \forall i \in \mathrm{bin}_k$.

\begin{align}
    &\int_{-\pi}^{\pi}\int_{-\pi}^{\pi} \cdots \int_{-\pi}^{\pi} \chi(\mathbf{c}) e^{-i \mathbf{c}.\mathbf{n}}\prod_{k=1}^{d^{(\mathrm{mb})}} \prod_{\{i,j\} \in \mathrm{bin}_k}\delta(c_i-c_j) {\prod_{j=1}^M} \frac{dc_j}{2\pi} \nonumber\\ 
    &=   \int_{-\pi}^{\pi} \cdots \int_{-\pi}^{\pi} \tilde{\chi}(\tilde{\mathbf{c}}) e^{-i \tilde{\mathbf{c}}.\tilde{\mathbf{n}}} {\prod_{k=1}^{d^{(\mathrm{mb})}}} \frac{d\tilde{c}_k}{2\pi} \\
    &= \int_{-\pi}^{\pi} \cdots \int_{-\pi}^{\pi} \sum\limits_{\bf n} P({\bf n}) e^{i \sum\limits_k \tilde{c}_k \sum\limits_{j\in \mathrm{bin}_k} n_j} e^{-i \tilde{\mathbf{c}}.\tilde{\mathbf{n}}} {\prod_{k=1}^{d^{(\mathrm{mb})}}} \frac{d\tilde{c}_k}{2\pi} \\
    &= \int_{-\pi}^{\pi} \cdots \int_{-\pi}^{\pi} \sum\limits_{\bf n} P({\bf n}) e^{-i \sum_k \tilde{c}_k(\tilde{n}_k - \sum\limits_{j\in \mathrm{bin}_k} n_j)}  {\prod_{k=1}^{d^{(\mathrm{mb})}}} \frac{d\tilde{c}_k}{2\pi} \\
    &= \sum\limits_{\bf n} P({\bf n}) \underbrace{ \int_{-\pi}^{\pi} \cdots \int_{-\pi}^{\pi} e^{-i \sum_k \tilde{c}_k(\tilde{n}_k - \sum\limits_{j\in \mathrm{bin}_k} n_j)}  {\prod_{k=1}^{d^{(\mathrm{mb})}}} \frac{d\tilde{c}_k}{2\pi}}_{=\prod\limits_{k=1}^{d^{(\mathrm{mb})}} \delta(\tilde{n}_k - \sum\limits_{j\in \mathrm{bin}_k} n_j)} \\
    &= \sum_{\bf n| \sum\limits_{j \in \mathrm{bin}_k} n_j = \tilde{n}_k}   P(\bf n) \\
    &=  P^{({\rm mb})}(\tilde{\bf n})
\end{align}

Note in the second step above we have defined
$\tilde{\chi}(\tilde{\bf c}) = \chi(\bf c)$, where $c_i = \tilde{c}_k \forall i \in \mathrm{bin}_k$.
We can replace the integral with a summation over grid points since we are always restricted to a fixed photon number space, in which case we arrive at 
\begin{align}
    P^{({\mathtt{mb}})}({\bf n})=\frac{1}{(N+1)^{d^{(\mathtt{mb})}}}\sum_{{\bf \tilde{c}}\in \mathbb{Z}_{N+1}^{d^{(\mathtt{mb})}}}\tilde{\chi}\left({\frac{2\pi \bf \tilde{c}}{N+1}}\right)e^{-i \frac{2\pi {\bf \tilde{c}}\cdot {\bf n}}{N+1} }.
\end{align}

\subsection{Payoff mechanism} \label{payoff}


To reward nodes for their work done in the boson-sampling subroutine, nodes are rewarded when their individual PBP $\mu_i$ is sufficiently close to the net PBP $\mu_{\mathtt{net}}$. That is, a reward $R_i = \mathcal{R}(\mu_i, \mu_{\mathtt{net}}, |s_i|)$ is paid to $node_i$ when $f(|\mu_i-\mu_{\mathtt{net}}|) < \epsilon$ is satisfied. To prevent cheating, a penalty term $P_i = \mathcal{P}(\mu_i, \mu_{\mathtt{net}}, |s_i|)$ is applied to $node_i$ when their individual PBP $\mu_i$ is far away compared to the net PBP $\mu_{\mathtt{net}}$ (i.e. $f(|\mu_i-\mu_{\mathtt{net}}|) \geq \epsilon$). The function $f$ should be monotonically increasing and we can assume it is linear in the argument.

We now construct a reward and penalty mechanism where it is the player's unique dominant strategy to behave honestly in the boson-sampling subroutine and not cheat. We construct $R_i$ and $P_i$ so that it scales linearly with the number of samples provided by $node_i$. Denote this as $n \equiv |s_i|$. We also denote R to be the base rate reward for satisfying $f(|\mu_i-\mu_{\mathtt{net}}|) < \epsilon$ with $n=1$ and let $P$ be the base rate penalty for satisfying $f(|\mu_i-\mu_{\mathtt{net}}|) \geq \epsilon$ with $n=1$. We also introduce a cutoff timestamp $T_{\mathtt{mine}}$ where only samples submitted before the cutoff time are considered for the payoffs. Finally, we denote the probability that an honest user satisfies the requirement $f(|\mu_i-\mu_{\mathtt{net}}|) < \epsilon$ as $p_i^{\mathtt{honest}}$ and the probability that a cheater satisfies the requirement $f(|\mu_i-\mu_{\mathtt{net}}|) < \epsilon$ as $p_i^{\mathtt{cheat}}$.

This gives the expected reward and payoff for $node_i$ as,
\begin{align}
  \mathbb{E}[R_i] &=
    \begin{cases}
      np_iR & \text{if $t_i < T_{\mathtt{mine}}$}\\
      0 & \text{otherwise}\\
    \end{cases}, \nonumber\\      
  \mathbb{E}[P_i] &=
    \begin{cases}
      n(1-p_i)P & \text{if $t_i < T_{\mathtt{mine}}$}\\
      0 & \text{otherwise}\\
    \end{cases},
\end{align}
where $p_i$ is either $p_i^{\mathtt{honest}}$ or $p_i^{\mathtt{cheat}}$ depending on the characteristics of $node_i$ as either a honest player or cheater. It is clearly sub-optimal to submit samples after the cutoff timestamp, thus the discussion going forward assumes that the player submits the samples prior to the cutoff time. There are 4 viable strategies for each player. They can:
\begin{itemize}
    \item Submit an honest sample from a quantum boson sampler (denoted with an ``honest'' superscript)
    \item Exit the PoW scheme and submit nothing (denoted with a ``nothing'' superscript)
    \item Submit a cheating sample from any algorithm (denoted with a ``cheat'' superscript)
    \item Submit an honest sample from a classical algorithm (denoted with a ``classical'' superscript)
\end{itemize}

We now show that given some innocuous assumptions, a payoff mechanism can be constructed such that a unique pure strategy Nash equilibrium exists where each player's dominant strategy is to submit an honest sample from a quantum boson sampler. To show this, we assume the following:
\begin{itemize}
    \item A player's utility is derived from the expected rewards minus the expected penalties and the costs incurred to generate a sample.
    \item An individual player's sample contribution is significantly smaller than the combined sample of all players (i.e. $|s_i| \ll |s_{total}|$) so that $\mu_{\mathtt{net}}$ remains unchanged irrespective of $node_i$ being honest or cheating.
    \item The verification subroutine is fairly accurate for $|s_i| \gg 1$ so that a honest player will satisfy $f(|\mu_i-\mu_{\mathtt{net}}|) < \epsilon$ with probability $p_i^{\mathtt{honest}} \in \mathbb{R}_{(0.75, 1)}$ and a cheater will satisfy $f(|\mu_i-\mu_{\mathtt{net}}|) < \epsilon$ with probability $p_i^{\mathtt{cheat}} \in \mathbb{R}_{(0, 0.25)}$.
    \item The cost to generate sample $\{s_{i}\}$ (denoted $C_i$) scales linearly with $|s_i|$. That is, $C_i = kn$, where $k \in \mathbb{R}$ and $n \equiv |s_i|$. The $k$ parameter includes costs such as energy consumption to generate one sample but should not include sunk costs \cite{Bernheim2014}. This assumption will be relaxed later to cover heterogeneous costs between players.
    \item The cost to generate a cheating sample is 0. This assumption will be relaxed later to cover cheating samples with costs.
\end{itemize}

We will cover the classical player later. Focusing on the first 3 strategies, the utilities are:
\begin{align}
    u_i^{\mathtt{honest}} &= \mathbb{E}[R_i] - C_i - \mathbb{E}[P_i] \nonumber \\
                 & = np_i^{\mathtt{honest}}R - nk - n(1-p_i^{\mathtt{honest}})P \nonumber \\
                 & = n(p_i^{\mathtt{honest}}R - k - (1-p_i^{\mathtt{honest}})P) &\nonumber \\
    u_i^{\mathtt{nothing}} &= 0 \nonumber \\
    u_i^{\mathtt{cheat}} &= \mathbb{E}[R_i] - \mathbb{E}[P_i] \nonumber \\
                 & = np_i^{\mathtt{cheat}}R - n(1-p_i^{\mathtt{cheat}})P \nonumber \\
                 & = n(p_i^{\mathtt{cheat}}R - (1-p_i^{\mathtt{cheat}})P)
\end{align}

To ensure that the dominant strategy is for players to behave honestly and for cheaters to exit the scheme we require that
\begin{align}
u_i^{\mathtt{honest}} > u_i^{\mathtt{nothing}} > u_i^{\mathtt{cheat}}.
\end{align}
So we require,
\begin{align}
    0 & < u_i^{\mathtt{honest}} \nonumber \\
    \implies 0 & < p_i^{\mathtt{honest}}R - k - (1-p_i^{\mathtt{honest}})P
    \nonumber \\
    0 & > u_i^{\mathtt{cheat}} \nonumber \\
    \implies 0 & > p_i^{\mathtt{cheat}}R - (1-p_i^{\mathtt{cheat}})P
\end{align}

Solving this, we obtain,
\begin{align}
\frac{p_i^{\mathtt{cheat}}R}{1-p_i^{\mathtt{cheat}}} < P < \frac{p_i^{\mathtt{honest}}R-k}{1-p_i^{\mathtt{honest}}} \label{eq:Pbound}
\end{align}

This inequality is not always defined. However, we note $p_i^{\mathtt{cheat}} < p_i^{\mathtt{honest}}$ and $\frac{1}{1-x}$ is increasing in $x \in \mathbb{R}_{(0,1)}$. So we have,
\begin{align}
\frac{1}{1-p_i^{\mathtt{cheat}}} < \frac{1}{1-p_i^{\mathtt{honest}}},
\end{align}
and a sufficient condition for inequality is,
\begin{align}
    p_i^{\mathtt{cheat}}R & < p_i^{\mathtt{honest}}R-k \nonumber \\
    \implies \frac{k}{p_i^{\mathtt{honest}}-p_i^{\mathtt{cheat}}} & < R.
\end{align}
Since,
\begin{align}
1 < \frac{1}{p_i^{\mathtt{honest}}-p_i^{\mathtt{cheat}}} < 2,
\end{align}
a sufficient condition for $R$ is,
\begin{align}
\frac{k}{p_i^{\mathtt{honest}}-p_i^{\mathtt{cheat}}} < 2k < R,
\end{align}
to ensure Eq.~\ref{eq:Pbound} is well-defined. Taking the tightest bounds for Eq.~\ref{eq:Pbound} and $2k < R$, we can bound $P$ by,
\begin{align}
\frac{1}{3}R < P < R.
\end{align}

These bounds ensure that,
\begin{align}
u_i^{\mathtt{honest}} > u_i^{\mathtt{nothing}} > u_i^{\mathtt{cheat}},
\end{align}
is satisfied and the dominant strategy for $node_i$ is to be honest.

\subsubsection{Classical Honest Players}

To keep the PoW protocol quantum and to disincentivize classical players from submitting samples to the network would require the utility of classical players to be negative while keeping the utility of quantum players positive. From the construction above, we have already derived bounds for $node_i$ to be honest. We will keep these bounds and derive an upper bound for $R$ that ensures $u_i^{\mathtt{honest}} > 0$ and $u_i^{\mathtt{classical}} < 0$.

We work under the assumption that the utility of a classical player is analogous to the utility of an honest player. That is,
\begin{align}
    u_i^{\mathtt{classical}} &= n(p_i^{\mathtt{classical}}R - k^{\mathtt{classical}} - (1-p_i^{\mathtt{classical}})P)
\end{align}

Where $p_i^{\mathtt{classical}} = p_i^{\mathtt{honest}}$ and $k^{\mathtt{classical}} \gg k$. It is reasonable to think of a classical player as performing the boson-sampling subroutine using a classical simulator instead of a true quantum boson-sampler. Letting $N$ be the number of photons and $M=N^2$ be the number of modes, the most efficient known classical boson-sampling simulator has a per sample cost proportional to the inverse of the repetition rate, $R_c$, defined in Eq.~\ref{Rclassical}, i.e.  $k^{\mathtt{classical}} \in O(2^N N)$. In contrast, a quantum boson sampler has a per-sample cost proportional to the inverse of the repetition rate $R_q$ (Eq.~\ref{Rquantum}). In the ideal case $(\eta_f=\eta_t=1)$, this cost is linear in $N$, otherwise, it increases exponentially with $N$ and $M$. However, as shown in Fig.~\ref{fig:analysis2} there is a large region of $N$values where this cost is several orders of magnitude smaller than that for classical supercomputers. Hence we can safely assume $k^{\mathtt{classical}} \gg k$.

To have $u_i^{\mathtt{classical}} < 0$, it is sufficient to have,
\begin{align}
k^{\mathtt{classical}} > R > p_i^{\mathtt{classical}}R,
\end{align}
since $p_i^{\mathtt{classical}} \in \mathbb{R}_{(0.75,1)}$. Combined with the derived bounds for $node_i$ to be honest, we have the bounds for $R$ and $P$ be,
\begin{align}
    2k & < R < k^{\mathtt{classical}}, \label{eq:RBound} \\
    \frac{1}{3}R & < P < R, \label{eq:Pbound2}
\end{align}

This ensures that $u_i^{\mathtt{honest}} > 0$, $u_i^{\mathtt{cheat}}$, $u_i^{\mathtt{classical}} < 0$, and $u_i^{\mathtt{nothing}} = 0$ and the dominant strategy of $node_i$ is to submit an honest sample to the network using a quantum boson-sampler. This strategy is unique as strictly dominant Nash equilibria are unique \cite{Tadelis2012}.

\subsubsection{Non-Nash Equilibrium without Penalty Term}

\cite{Kroll2013} showed that under certain assumptions, deterministic tests to check PoW can have a Nash equilibrium that is in line with the consensus protocol's best interests. In this section, we show that contrary to deterministic tests to check PoW (such as running double SHA-256 in Bitcoin), a penalty term is a necessity for statistical tests that check PoW to ensure it is a Nash equilibrium for players to remain honest. This is because statistical tests imply a non-zero probability of passing the test even though a player may have submitted a cheating sample. A penalty term ensures that it is not optimal for the cheater to submit cheating samples in this manner.

Without a penalty term, the utilities of the players are:
\begin{align}
    u_i^{\mathtt{honest}} &= \mathbb{E}[R_i] - C_i \nonumber\\
                 &= np_i^{\mathtt{honest}}R - nk \nonumber\\
                 &= n(p_i^{\mathtt{honest}}R - k), \nonumber\\
    u_i^{\mathtt{nothing}} &= 0 \nonumber\\
    u_i^{\mathtt{cheat}} &= \mathbb{E}[R_i] \nonumber\\
                 &= Np_i^{\mathtt{cheat}}R,
\end{align}
where $n=|s_i^{\mathtt{honest}}|$ is the number of samples committed by an honest player and $N=|s_i^{\mathtt{cheat}}|$ is the number of samples committed by a cheater. To show that the honest strategy is not a Nash equilibrium, it suffices to show that $u_i^{\mathtt{cheat}} > u_i^{\mathtt{honest}}$. Let $N=\frac{np_i^{\mathtt{honest}}}{p_i^{\mathtt{cheat}}}$. Then,
\begin{align}
    u_i^{\mathtt{cheat}} & = Np_i^{\mathtt{cheat}}R \nonumber\\
    &= \frac{np_i^{\mathtt{honest}}}{p_i^{\mathtt{cheat}}}p_i^{\mathtt{cheat}}R \nonumber\\
    &= np_i^{\mathtt{honest}}R \nonumber\\
    &> n(p_i^{\mathtt{honest}}R-k) \nonumber\\
    &= u_i^{\mathtt{honest}}.
\end{align}

In essence, when sample submission incurs negligible costs (i.e. $k=0$) and without a penalty term, cheaters could artificially inflate their sample size in hopes of getting a large payoff by chance. This would result in a higher utility to act maliciously and destroy the original Nash equilibrium of being honest.

\subsubsection{Heterogeneous Costs}

We now relax the assumption that all players have the same cost factor $k$ for generating one sample by a quantum boson-sampler and allow for a heterogeneous cost factor. That is, for player $i \in \{1,2,...,p \}$ with cost function $C_i = k_in$, $k_i \in \mathbb{R}_{>0}$ is potentially different along the players.

With heterogeneous costs, we set the cost factor $k$ in Eq.~\ref{eq:RBound} to the cost factor of the most efficient player (i.e. $k=\min\{k_1, k_2,...,k_p \}$). This ensures that there is at least one player (the most efficient player) such that,
\begin{align}
u_{eff}^{\mathtt{honest}} > u_{eff}^{\mathtt{nothing}} > u_{eff}^{\mathtt{cheat}}.
\end{align}
Since the sign of $u_i^{\mathtt{cheat}}$ is independent of the value of $k$, this also ensures that $u_i^{\mathtt{cheat}} < u_i^{\mathtt{nothing}}=0$ for $i \in \{1,2,...,p\}$. For inefficient players with individual cost factors $k_i > k$ such that $u_i^{\mathtt{honest}} < u_i^{\mathtt{nothing}}$, the market mechanism will have the inefficient players leave the PoW scheme and submit nothing for verification.

If the variation in the individual cost factors is significant enough such that setting $k$ to be the most efficient cost factor will result in the market becoming saturated, we can set $k$ to be the $m$th lower percentile cost factor (i.e. $k=\min_{m\%}\{k_1, k_2,...,k_p \}$) so that we can ensure at least $m$ per cent of the $p$ players will have a positive payoff from contributing samples to the network and not exit the PoW scheme.

\subsubsection{Cheating with Costs}

If players have non-zero costs for generating a cheating sample, then it is clearly sub-optimal for players to cheat since cheating without costs is already a dominant strategy. Additional costs associated with cheating just make the utility for cheaters lower.

\subsubsection{Block Reward vs. Split Reward}

The derivations above assumed a split reward mechanism. That is, the reward for the addition of a new block is split between all players satisfying $f(|\mu_i-\mu_{\mathtt{net}}|) < \epsilon$ and each player receives $nR$ for their $n$ samples provided. Another reward mechanism that could be used is a block reward mechanism in which the entire reward is awarded to one player instead of splitting it between players (i.e. one player satisfying $f(|\mu_i-\mu_{\mathtt{net}}|) < \epsilon$ would randomly be chosen to receive the entire reward). While the expected reward would stay the same, there is now considerable variation in the payoff for the player. The initial assumption that the player's utility is risk-neutral and only depends on the expected rewards/penalties and costs would no longer be valid.

Conventional mean-variance utility theory in finance imposes a penalty term for risk-aversion due to the variability of the payoffs \cite{Markowitz1952, Bailey2005}. Thus, for block reward mechanisms, it is more appropriate to use utility functions of the form
\begin{align}
    u_i &= \mathbb{E}[R_i] - C_i - \mathbb{E}[P_i] - A_i\sigma^2.
\end{align}

Where $A_i$ is the coefficient of risk-aversion for $node_i$ and $\sigma^2$ is the variance of the reward. It is difficult if not impossible to estimate the parameter $A_i$ for all the players in the PoW protocol as it is intrinsically related to individual preferences of risk-aversion. We do not claim here that we can provide an estimate, empirical or theoretical, for its value. However, for implementation purposes, the reward $R$ in Eq.~\ref{eq:RBound} should be set higher for a block reward mechanism compared to a split reward mechanism so that the additional expected rewards $\mathbb{E}[R_i]$ would offset the penalty from risk-aversion $A_i\sigma^2$.

For implementation purposes of a block reward mechanism, it may also be prudent to consider safeguards against selfish miners as proposed in \cite{EyalandSirer2013}. In their paper, the authors discussed a mining protocol that deviated from the intended consensus protocol and had revenues scaling super-linearly against computational power once a threshold percentage of the network is dominated by one party (the authors upper bound this threshold by 1/3). This is particularly relevant to block reward mechanisms due to the formation of mining pools to reduce the variance of payoffs. As such, it may be prudent to implement the solution proposed in \cite{EyalandSirer2013} that raises the threshold to 1/4. That is, whenever the blockchain forks and two branches of length one occur, instead of each node mining the branch that they received first, the protocol should dictate that they randomly and uniformly choose one of the two branches to mine. The act of this randomization safeguards against potential selfish miners that control less than 1/4 of the computational power of the network.

\subsubsection{Components of Costs (Variable $k$) and Cost to Entry}

The cost variable $k$ (or $k_i$ for heterogeneous costs) is the amalgamation of all relevant costs to the generation of one sample. There is a distinction in this cost factor for players wishing to enter the boson-sampling scheme (prospective players) and for players already providing samples to the boson-sampling subroutine (current players).

For current players or players using a subscription-based cloud boson sampler, the cost factor $k$ should only include the variable costs required to produce one sample to the sampling subroutine (e.g. subscription costs, electricity costs, boson preparation costs, measurement costs). That is, $k = k_{variable}$. The fixed cost of the boson-sampling device is sunk and its cost should not be taken into consideration for sampling decisions going forward \cite{Bernheim2014}.

For prospective players, however, the initial capital expenditure costs (e.g. source guides, detectors, machinery) must be taken into consideration for $k$. If $\tau$ is the expected number of samples the boson-sampler is expected to produce before obsolescence, then,
\begin{align}
k = k_{variable} + \frac{k_{fixed}}{\tau}.
\end{align}
For the PoW protocol to be self-sustaining in the long run with consistent user renewal, the value for $k$ in Eq.~\ref{eq:RBound} must be above the $k$ value for prospective players so that there are sufficient incentives for new players to overcome the cost to entry.

Two comments are worth adding here on the adoption of this new PoW consensus protocol. First, in the early stages, before large-scale production and availability of boson samplers, it could be expected that classical miners would dominate. This could be accommodated by having the reward inequality in Eq.~\ref{eq:RBound} be initially $R > k^{\mathtt{classical}}$ so that the utility of classical players is positive. Then a decision could be made to gradually either (1) increase $k^{\mathtt{classical}}$ (such as increasing the number of photons in the sampling problem and hence the difficulty) or (2) reduce $R$. This will kick classical players out of the protocol as they no longer have positive utility.
Second, the conditions on reward and penalty described above assume that the Nash equilibrium is already reached since it is defined by the condition that no unilateral deviation will move the equilibrium. This will not be the case during the initialization stage of the protocol. During the genesis block and several blocks thereafter, additional mechanisms should be placed by trustworthy players, to ensure the initialization reaches this Nash equilibrium. The trustworthy players can then exit the market and the equilibrium will be retained, thus ensuring no ``central authority'' exists in the protocol.

\subsection{Parameters and description}

\begin{table*}[!htp] 
  \centering
  \begin{tabular}{|c|c|p{9.4cm}|p{3.7cm}|}
    \hline
    \multirow{1}{*}{} & Notation & Description & Comments \\
    \hline
    \multirow{2}{*}{BS setup} & $N$ & Total number of input photons &  \\
  & $M$ & Total number of modes & We use $M=O(N^2)$ \\
   & $U$ & Matrix description of linear optical circuit & $U\in \mathbb{C}^{M\times M}$ \\
    \hline
    \multirow{7}{*}{\parbox{2cm}{State-binned BS}}  
    & $d^{(\mathtt{sb})}$ & Number of bins in the state Hilbert space & $d^{(\mathtt{sb})}\lesssim 0.1/2\epsilon^{0.8}$ \\
    & $\mu_{\mathtt{true}}$ & Peak bin probability (PBP) of state-binned BS & \\
    & $\mu_{\mathtt{net}}$ & Estimated PBP from all validated samples on the network & \\
     & $\epsilon$ & Desired accuracy of the estimated PBP wrt. $\mu_{\mathtt{true}}$ &  \\
       & $N^{(\mathtt{sb})}_{tot}$ & The number of samples needed across the network such that \mbox{$|\mu_{\mathtt{net}}-\mu_{\mathtt{true}}|\leq \epsilon$} with high confidence & $N^{(\mathtt{sb})}_{\rm tot}= 1.8\times 10^5 {d^{(\mathtt{sb})}}^{7/2}$ \\
       & $\gamma$ & $100(1 - \gamma)\%$ is the confidence interval for $\mu_{\mathtt{true}}$ & We assume $\gamma=10^{-4}$ \\

    \hline
    \multirow{15}{*}{\parbox{2cm}{Mode-binned BS}} & $d^{(\mathtt{mb})}$ & Number of bins in the mode Hilbert space & \\
         & $P^{({\mathtt mb})}[i]$ & Mode-binned probability distribution for miner $i$ & \\
      & $m$ & Number of random samples needed to estimate matrix permanent up to an $\delta$ additive estimate with probability at least $p$, using Gurvits' algorithm & $m=\frac{2}{\delta^2}\ln(2/(1-p))$ \\
       & $\beta$ & Accuracy parameter used to invalidate a miner's mode binned distribution based on total variation distance from an estimated mode binned distribution $\widehat{P^{({\mathtt{mb}})}}$ calculated using Gurvits' algorithm &   \mbox{$\mathcal{D}^{({\rm tv})}(\widehat{P^{({\mathtt{mb}})}},P^{({\mathtt{mb}})}[i])\geq 2\beta$} \\
        & $N^{(\mathtt{mb})}_{tot}$ & The number of samples sufficient for a quantum boson sampler to pass the validation test & $N^{(\mathtt{mb})}_{{\rm tot}}=2^{14}\frac{\sqrt{\binom{N+d^{(\mathtt{mb})}-1}{N}}}{\beta^2}$\\                            
          & $m_{j}[i]$ & Number of photons in the $j^{th}$ bin for the $i^{th}$ sample set $s_{i}$ & \\  
                            & $s_{i}^{(v)}$ & Set of samples committed by the $i^{th}$ node in the blockchain network. 
                             & The $v$ superscript, if any, denotes the samples are validated \\       & $W^{(v)}$ & The combined set of samples committed by all the nodes in the blockchain network & $W^{(v)}=\bigcup_i s^{(v)}_i$.  \\
    \hline
    \multirow{7}{*}{\parbox{2cm}{Experimental resources}} 

                    & $\eta_{f}$ & Combined parameter for single photon generation, coupling, and detection efficiency & \\
                    & $\eta_{t}$ & Transmission probability of a single beam-splitter & \\
                    & $R_{0}$ & Single photon repetition rate & \\
                    & $R_{q}$ & Repetition rate of a quantum boson sampler & $R_q=(\eta_f\eta^M_t)^N R_0/N e$\\
                    
    \hline
    \multirow{7}{*}{\parbox{2cm}{Payoff mechanism}}
                    & $\mu_{i}$ & Peak bin  probability over the $i^{th}$ verified sample set $s_{i}^{v}$ & \\
                    & $t_{i}$ & Time when the $i^{th}$ user commits their samples to the network & \\
                    & $p_{i}^{\mathtt{honest}}$ & Probability of an honest user (either classical or quantum sampler) to pass the 
                    mode-binned BS test & \\
                    & $p_{i}^{\mathtt{cheat}}$ & Probability of a user providing non-BS samples to pass the mode-binned BS test & \\
                    & $u_{i}^{\mathtt{honest}}$ & Utility of an honest node in the network & \\
                    & $u_{i}^{\mathtt{cheat}}$ & Utility of a cheating node in the network & \\
                    & $u_{i}^{\mathtt{nothing}}$ & Utility of doing nothing, i.e. not committing any samples & \\
                    & $k^{\mathtt{classical}}$ & Cost per honest sample using classical algorithms & \\
                    & $k^{\mathtt{quantum}} \equiv k$ & Cost per honest sample using actual BS implementation & \\
                    & $C_{i}$ & The cost incurred by the $i^{th}$ node in committing $|s_{i}|$ samples & \\
                    & $P_{i}$ & Penalty implemented on the $i^{th}$ node in the network & \\
                    & $R_{i}$ & Reward awarded to the $i^{th}$ node in the network & \\
    \hline
  \end{tabular}
    \caption{List of parameters used in different modules of the algorithm.}
    \label{tab:grouped-rows}
\end{table*}

\end{document}